\documentclass[aps, prl, 10pt, twocolumn, superscriptaddress, preprintnumbers, floatfix]{revtex4-2}

\usepackage[colorlinks=true,breaklinks=true]{hyperref}
\hypersetup{allcolors=[rgb]{0.0 0.0 0.75},linkcolor=[rgb]{0.75 0.05 0.05}}
\usepackage[dvipsnames]{xcolor}
\usepackage{mathtools}
\usepackage{amsmath}
\usepackage{amsfonts}
\usepackage{amssymb}
\usepackage{bbold}
\usepackage{upgreek}
\usepackage{bm}
\usepackage{soul}
\long\def\exclude#1{}
\usepackage{orcidlink}
\usepackage{slashed}
\usepackage{microtype}

\newcommand{\bq}{{\bf q}}

\begin{document}

\title{Neutron Star Bounds on Muonic Fifth Forces from Picometer to Meter Scales}

\author{Damiano F.\ G.\ Fiorillo \orcidlink{0000-0003-4927-9850}}
\affiliation{Istituto Nazionale di Fisica Nucleare (INFN), Sezione di Napoli, Complesso Universitario di Monte Sant'Angelo, Via Cintia, 80126 Napoli, Italy}
\affiliation{Gran Sasso Science Institute (GSSI), Viale Francesco Crispi 7, 67100 L’Aquila, Italy}

\author{Alessandro~Lella~\orcidlink{0000-0002-3266-3154}}
\affiliation{Dipartimento di Fisica e Astronomia, Università degli Studi di Padova, Via Marzolo 8, 35131 Padova, Italy}
\affiliation{Istituto Nazionale di Fisica Nucleare (INFN), Sezione di Padova, Via Marzolo 8, 35131 Padova, Italy}%

\author{Georg~G.~Raffelt~\orcidlink{0000-0002-0199-9560}}
\affiliation{Max-Planck-Institut f\"ur Physik, Boltzmannstra\ss e~8, 85748 Garching, Germany}

\author{Nud{\v z}eim~Selimovi{\'c}~\orcidlink{0000-0003-3780-1437}}
\affiliation{Dipartimento di Fisica e Astronomia, Università degli Studi di Padova, Via Marzolo 8, 35131 Padova, Italy}
\affiliation{Istituto Nazionale di Fisica Nucleare (INFN), Sezione di Padova, Via Marzolo 8, 35131 Padova, Italy}%

\author{Edoardo~Vitagliano~\orcidlink{0000-0001-7847-1281}}
\affiliation{Dipartimento di Fisica e Astronomia, Università degli Studi di Padova, Via Marzolo 8, 35131 Padova, Italy}
\affiliation{Istituto Nazionale di Fisica Nucleare (INFN), Sezione di Padova, Via Marzolo 8, 35131 Padova, Italy}

\begin{abstract}
Testing muonic fifth forces is difficult because muons are scarce in ordinary matter. However, they are abundant in neutron stars. We show that this population responds coherently to electromagnetic fluctuations induced by proton and electron collisions, emitting the fifth-force mediator and cooling the star. This energy loss constrains $g_{X\mu}\alt 10^{-13}$ for scalars with $m_\phi\alt 100$~keV and vectors with $10^{-5}\,\mathrm{eV}\alt m_V\alt 100$~keV, far more strongly than SN~1987A cooling. Vector forces at smaller $m_V$ deplete muons, suppressing the coherent response. Cooling from incoherent muon scattering still constrains $g_{V\!\mu}\alt 2.5\times 10^{-12}$ for $1.3\times 10^{-7}\,\mathrm{eV}\alt m_V\alt 10^{-5}$~eV. Together, these bounds provide the most sensitive probes of muonic interactions across a broad range of parameters.
\end{abstract}

\maketitle

\textbf{\textit{Introduction}}---New light bosons could mediate forces beyond the four known fundamental interactions. Casimir measurements~\cite{Sushkov:2011md, Chen:2014oda}, microcantilevers~\cite{Geraci:2008hb}, torsion-balance experiments~\cite{Kapner:2006si, Lee:2020zjt, Smith:1999cr, Yang:2012zzb, Tan:2020vpf, Hoskins:1985tn}, and satellite-borne accelerometers~\cite{Berge:2017ovy} provide powerful probes of fifth forces. As they rely on ordinary matter, however, they have limited sensitivity to interactions coupling primarily to muons.

Muonic bosons have been proposed as gauge bosons of a new $U(1)_{L_\mu-L_\tau}$ symmetry~\cite{Foot:1990mn, He:1990pn, He:1991qd, Foot:1994vd}, as scalars~\cite{Batell:2017kty, Batell:2021xsi}, or as pseudoscalars (e.g.\ Ref.~\cite{DiLuzio:2020wdo} and references therein). They are phenomenologically described by a gauge or Yukawa coupling $g_{X\!\mu}$ and mass~$m_X$, corresponding to an interaction range $\lambda_X \sim m_X^{-1}$, where $X=\phi,a,V$ denotes scalar, pseudoscalar, and vector bosons. The quest for muonic fifth forces then amounts to the question: how strongly can these bosons couple to muons?

Remarkably, neutron stars (NSs) provide particularly powerful tests. In these compact objects, muons are abundant despite their large mass, $m_\mu=105.7$~MeV, owing to beta equilibrium. In a reference model for a middle-aged NS (see End Matter), $Y_p\simeq0.11$ is a representative proton fraction per baryon, with $Y_e\simeq0.07$ and $Y_\mu\simeq0.04$. Previous studies used the large muon densities of pulsars \cite{KumarPoddar:2019ceq, Dror:2019uea} and proto-neutron stars \hbox{\cite{Bollig:2020xdr, Croon:2020lrf, Caputo:2021rux, Akita:2023iwq, Blinov:2025aha, Ferreira:2025qui, Das:2026uam}} to constrain forces of very long and very short range. 

\begin{figure*}
  \centering
  \hbox to \textwidth{\includegraphics[width=0.5\textwidth]{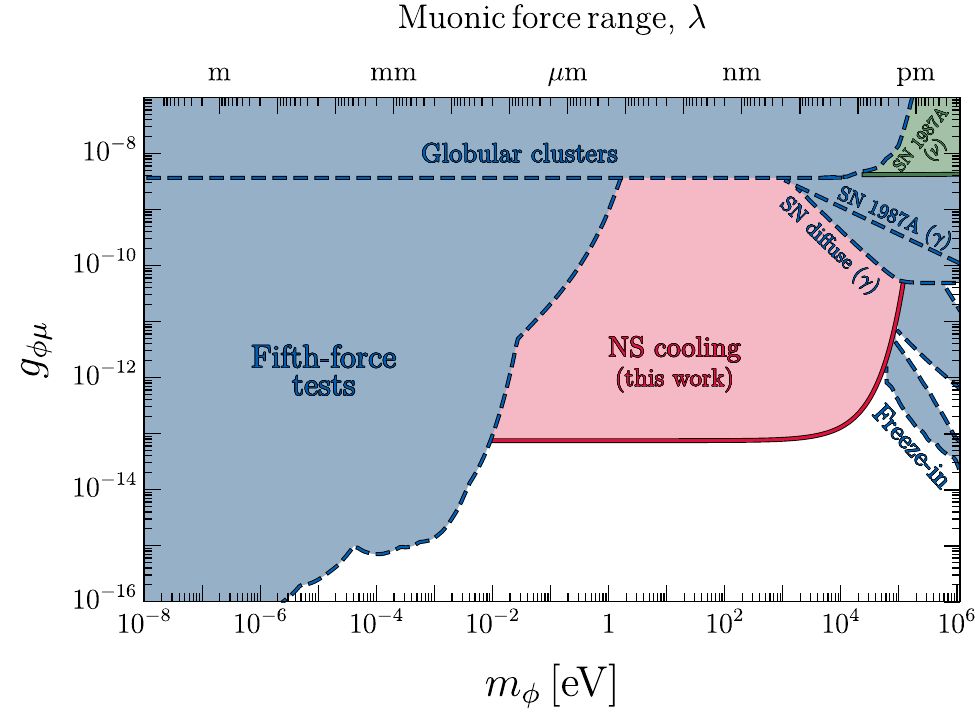}\hfill
  \includegraphics[width=0.5\textwidth]{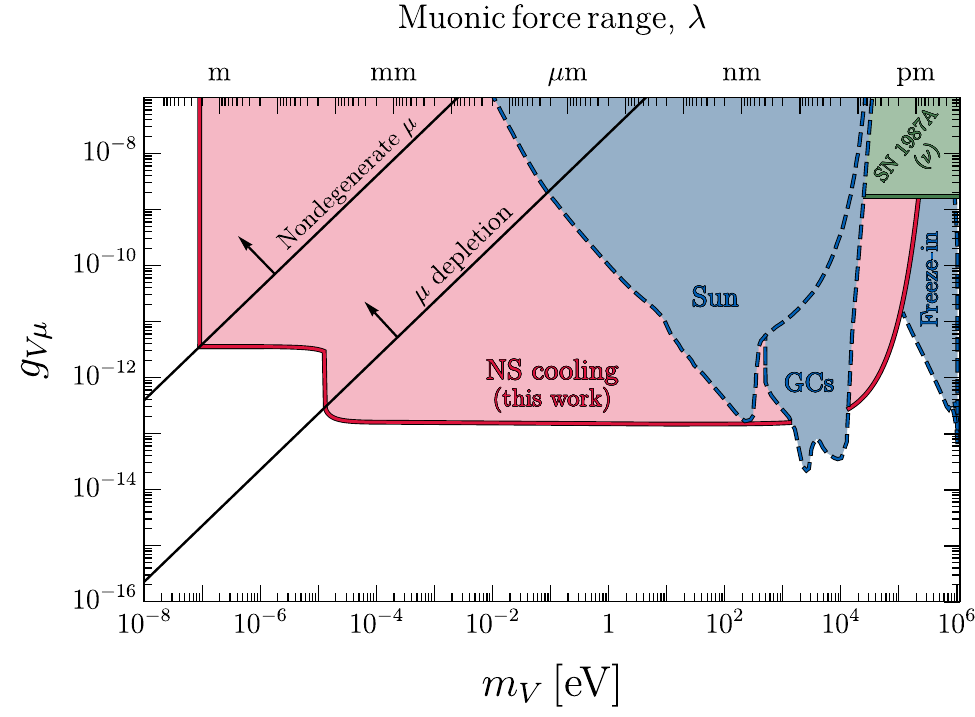}}
  \vskip-4pt
  \caption{Bounds on the coupling of muonic scalars (\emph{left}) and vectors (\emph{right}) as a function of mass $m_X$, or force range $\lambda=1/m_X$ (\emph{upper axis}). Solid lines show tree-level bounds, dashed lines loop-level ones. The supernova (SN) and globular cluster (GC) bounds on scalars are from Ref.~\cite{Caputo:2021rux} with updates from \cite{Dolan:2022kul}, while those from an irreducible cosmological scalar population are from~\cite{Langhoff:2022bij}. Fifth-force bounds~\cite{Hoskins:1985tn, Smith:1999cr, Kapner:2006si, Geraci:2008hb, Sushkov:2011md, Yang:2012zzb, Chen:2014oda, Decca:2007yb, Blakemore:2021zna, Venugopalan:2024kgu, Chiu:2009fqu, Bezerra:2010pq, Banishev:2012kkb, Banishev:2014jka, Berge:2017ovy, Lee:2020zjt, Tan:2020vpf, Matias:2025vrq, Boynewicz:2025daj, Ren:2026hff} come from loop-induced couplings to protons (see~text). For vectors, the oblique black lines identify the regions where long-range forces significantly reduce the muon abundance~(\emph{$\mu$ depletion}), eventually making muons nondegenerate~(\emph{nondegenerate $\mu$}); see text for details. The SN~1987A cooling bound is from Ref.~\cite{Blinov:2025aha}. Additional constraints (blue) assume that the vector couples to tau as well. They are obtained by rescaling dark-photon kinetic-mixing  bounds with $\epsilon=-g_{V\!\mu}/70$. Solar and GC bounds are from Refs.~\cite{Li:2023vpv,Dolan:2023cjs}. Further constraints on vector bosons arise from a cosmological population produced through freeze-in, leaving imprints in galactic and diffuse X-ray observations~\cite{Cirelli:2010xx, Linden:2024fby, Caputo:2025avc}.}
  \label{fig:Bounds}
\end{figure*}

In this Letter, we show that long-term NS cooling is a sensitive probe of the intermediate regime for scalars and vectors with $10^{-7}\,\mathrm{eV}\lesssim m_X\lesssim 10^5\,\rm eV$ (Table~\ref{tab:Bounds} and Fig.~\ref{fig:Bounds}). The emission of these bosons scales approximately as $T^4$ with the internal NS temperature, so at late times it becomes increasingly important relative to standard $T^8$ neutrino cooling \cite{Friman:1979ecl, Bottaro:2024ugp}. In analogy to the baryophilic case \cite{Ishizuka:1989ts, Fiorillo:2025zzx}, this makes middle-aged NS cooling more sensitive than SN~1987A.

Despite this analogy, the underlying emission mechanism is qualitatively new. Over a large region of parameter space, the main emission is indirect, through the coherent response of the degenerate muons to electromagnetic fluctuations induced by proton and electron collisions. Moreover, for a sufficiently long-range vector force, the backreaction on muons suppresses their abundance, eliminating the constraint when $\lambda_V\agt 1\,\rm m$.

\begin{table}[t!]
\vskip-4pt
 \caption{Stellar cooling bounds on the gauge or Yukawa coupling $g$ of muonic bosons ($m_X\alt 1$~keV). The globular cluster (GC) bound derives from $G_{X\gamma\gamma}$ through a muon loop.
  \label{tab:Bounds}}
 \vskip4pt
    \begin{tabular*}{\columnwidth}{@{\extracolsep{\fill}}llll}
    \hline\hline
&Pseudo&Scalar&Vector\\
\hline
NS cooling (new)& $2.9\times10^{-9}$ & $7.4\times10^{-14}$ & $1.5\times10^{-13}$\\
SN~1987A cooling \cite{Caputo:2021rux}$^a$& 
$9.1\times10^{-9}$&$4.2\times10^{-9}$&$2.7\times10^{-9}$\\
GC stars \cite{Caputo:2021rux}$^b$&
$2.6\times10^{-9}$&$3.9\times10^{-9}$&---\\
\hline
\end{tabular*}
\vskip4pt
\hbox to\columnwidth{$^a$Conservative bounds from the cold SN model.\hfill}
\hbox to\columnwidth{$^b$Updated to new ALP bound from 
GC stars \cite{Dolan:2022kul}.
\hfill}
\end{table}

{\bf\textit{Essentials of NS cooling}}---Following core collapse, a NS cools via neutrino emission for about $10^5$~years, after which photons take over. The mechanical structure is set by degenerate matter, largely decoupled from the thermal evolution set by $T$ of the isothermal core. The heat capacity scales as~$T$, so the internal energy is $U=C_U T^2$. The neutrino luminosity scales as $L_\nu=C_\nu T^8$, implying $T\propto t^{-1/6}$. The photon luminosity is $L_\gamma \propto T_{\rm s}^4$, where the surface temperature satisfies $T_{\rm s}\propto T^{1/2}$ because the conductivity of degenerate heat carriers scales~as~$T$. Hence, $L_\gamma=C_\gamma T^2\propto t^{-1/3}$ during neutrino cooling. When photon cooling takes over, the common scaling of $U$ and $L_\gamma$ with $T^2$ implies $L_\gamma\propto e^{-t/\tau}$ with $\tau\simeq10^6$~years.

\begin{figure}
  \centering
  \includegraphics[width=\columnwidth]{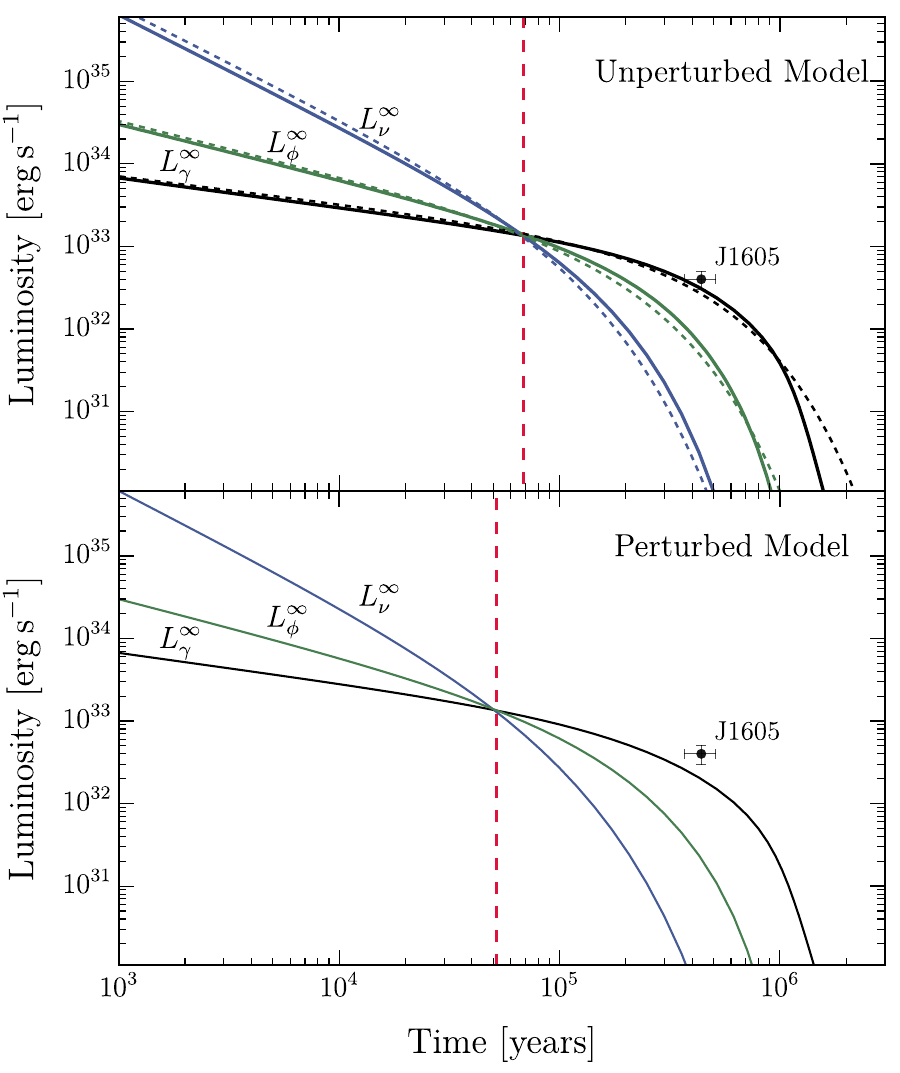}
  \caption{Evolution of $L_\gamma$ (black), $L_\nu$ (blue) and $L_\phi$ (green) in our reference NS model simulated with \texttt{NSCool}, where $\phi$ is a baryophilic scalar with $g_{\phi N}\simeq9.5\times10^{-14}$, saturating the NS cooling criterion (see text). The observed NS is J1605. \emph{Upper panel:} Unperturbed model, dashed lines showing the analytic scalings of $L_\gamma$, $L_\nu$ and $L_\phi$. Here, $L_\phi$ derives from post-processing the unperturbed model. \emph{Lower panel:} $L_\phi$ is self-consistently included in {\tt NSCool} as a new cooling channel. The red dashed lines indicate the epochs when the different cooling channels become equal, which occurs at $T_{\rm core}\simeq 8.9\,$keV.
  }\label{fig:Cooling}
\end{figure}

This simple picture agrees with a numerical reference model detailed in the End Matter and shown in Fig.~\ref{fig:Cooling}. The analytic lines (dashed) use $C_\nu= 1.06$ and $C_\gamma=1.27$ (luminosities in $10^{33}~{\rm erg}~{\rm s }^{-1}$, $T$ in $10^8$~K), and $\tau=7\times10^{5}$~years, chosen to match the numerical curves (solid).
Neutrino-photon equality occurs at $t\simeq7\times10^{4}$~years with $T\simeq8.9\,{\rm keV}$ and $L_\nu=L_\gamma\simeq 1.4\times 10^{33}~{\rm erg}~{\rm s}^{-1}$.

Realistic NS evolution is complicated by uncertainties of the nuclear Equation of State~(EoS), nucleon superfluidity or other novel phases, neutrino emissivities, strong magnetic fields, and overall stellar parameters. Despite these complications, Refs.\ \cite{Buschmann:2021juv, Fiorillo:2025zzx} constrained new cooling channels by comparing measured ages and luminosities of several isolated middle-aged NSs with {\tt NSCool} \cite{2016ascl.soft09009P} models. This public code allows one to switch between different EoSs and different assumptions about neutron and proton superfluidity. New cooling channels were then constrained by varying the NS mass and marginalizing over these nuisance parameters.

For both axions \cite{Buschmann:2021juv} and baryophilic bosons \cite{Fiorillo:2025zzx}, the resulting constraint is driven by the pulsar J1605, whose age and luminosity are shown in Fig.~\ref{fig:Cooling} (see~\cite{Tetzlaff:2012rz,Pires:2019qsk}). Using the standard EoS APR~\cite{Akmal:1998cf} without superfluidity, the best-fit baryonic mass of the unperturbed model is $1.8\,M_\odot$, which we use as our reference NS; more detailed properties are shown in the End Matter. The improved neutrino rates of Bottaro et al.\ \cite{Bottaro:2024ugp} instead of those of Friman and Maxwell~\cite{Friman:1979ecl} further improve the agreement of the standard cooling curve in Fig.~\ref{fig:Cooling} with J1605.

Since muonic bosons affect NS cooling similarly to baryophilic bosons, we adopt a simplified criterion rather than repeating the full analysis.

{\bf\textit{Particle bounds}}---A new energy-loss channel, $L_X=C_X T^{p}$, modifies standard evolution. Figure~\ref{fig:Cooling} illustrates the case $p=4$, relevant for baryophilic scalars \cite{Ishizuka:1989ts, Fiorillo:2025zzx} and muonic scalars and vectors. We choose $C_\phi$ so that, in the unperturbed model, $L_\phi=L_\nu=L_\gamma$ at neutrino–photon equality. Thin solid curves show the evolution when $L_\phi$ is included self-consistently. Since $L_\nu\propto T^8$ and $L_\gamma\propto T^2$, a channel with $2<p<8$ has its greatest impact around this epoch and simply shortens the cooling time.

For baryophilic scalars, the full statistical analysis yields $g_{N}\lesssim 5\times10^{-14}$ at 95\%~CL \cite{Fiorillo:2025zzx}. At this coupling, our reference star has $L_\phi=3.7\times10^{32}\,{\rm erg\,s^{-1}}$ at neutrino-photon equality. Thus, the limit is reproduced by the simple requirement $L_\phi\lesssim 0.3\,L_\nu$ at $T=8.9$~keV. 

In other words, the intuitive requirement $L_X\lesssim \,L_\nu$, evaluated at neutrino-photon equality, provides a conservative criterion; the modest gain of a detailed analysis lies well within astrophysical uncertainties. We therefore adopt this condition to constrain muonic bosons. Equivalently, for an average baryonic density of $8.5 \times 10^{14}~{\rm g}~{\rm cm}^{-3}$, the new bosons should obey
\begin{equation}
\epsilon_X\lesssim 0.3\,{\rm erg}~{\rm g}^{-1}~{\rm s}^{-1}
\end{equation}
to avoid excessive cooling. This approximate bound depends only weakly on the power-law index $p$ and essentially coincides with Iwamoto's original criterion \cite{Iwamoto:1984ir}.

{\bf\textit{Muonic boson emission}}---Calculating the emission rate is surprisingly challenging. For pseudoscalars, muonic bremsstrahlung is the main source, with the dominant electromagnetic exchange coming from Landau-damped transverse modes rather than Debye-screened longitudinal ones. For scalars and vectors, the main emission instead comes from coherent fluctuations of the muon gas, induced primarily by electron and proton scattering, respectively. In other words, medium effects generate effective couplings of muonic bosons to electrons and protons. We defer a detailed study to our technical companion paper ~\cite{Fiorillo:2026long}  and here only provide intuitive scaling arguments.

To set the stage, we begin with the simple process of vector bremsstrahlung, $\mu\mu\to\mu\mu V$. Since the boson frequency $\omega\sim T$ is small, this is a classical radiation process. In a typical collision, the muon velocity changes by $\Delta v=q/\mu_\mu$, where $q$ is the exchanged momentum and $\mu_\mu$ the Fermi energy. For mildly relativistic muons, transverse  electromagnetic exchange dominates because longitudinal fields are strongly screened. Under superconducting conditions, the screening distance of currents is set by the Meissner scale $m_M$, so $q\sim m_M$. Since the dipole radiation probability scales as $g_{V\mu}^2 \Delta v^2$ \cite{Landau:1975pou}, and each quantum carries energy of order $T$, we find 
\begin{equation}\label{eq:QV_mumu_generic}
    Q_V^{(\mu\mu)}\sim \frac{g_{V\!\mu}^2 T m_M^2}{\mu_\mu^2}n_\mu^2 \sigma_{\mu\mu}F_{\rm deg},
\end{equation}
where $n_\mu\sim \mu_\mu^3$ and $\sigma_{\mu\mu}\sim \alpha^2/m_M^2$. 

The degeneracy suppression $F_{\rm deg}$ reflects the restricted muon phase space near the Fermi surface. Of the four momenta entering the collision, one is fixed by energy-momentum conservation, while the remaining three are restricted to widths of order $T$. In the nondegenerate case, the corresponding scales are $\mu_\mu$ for the two incoming particles and a change $q\sim m_M$ for the outgoing one. Hence, $F_{\rm deg}\sim T^3/\mu_\mu^2 m_M$, and therefore
\begin{equation}
    Q_V^{(\mu\mu)}\sim \frac{g_{V\!\mu}^2\alpha^2 \mu_\mu^2T^4}{m_M}.
    \label{Eq:QV_mumu}
\end{equation}
Without superconductivity, magnetic fields are not Meissner screened but Landau damped. Therefore,  $m_\Lambda\sim (\alpha \mu_\mu^2 T)^{1/3}$ replaces $m_M$ so that $Q_V^{(\mu\mu)}$ scales as $T^{11/3}$.

The dominant channel for vector emission, however, is the coherent muon response to rapid proton-neutron scattering. The proton current $\mathbf{j}_p$ creates a transverse electromagnetic field. The relevant frequencies are of order $T$, far below the plasma frequency, so this field is instantly screened by a medium response $\mathbf{j}_{\rm med}=-\mathbf{j}_p$. If $\Pi_{T,\alpha}$ denotes the transverse susceptibility of species $\alpha=e,p,\mu$ with $\Pi_T=\sum_\alpha \Pi_{T,\alpha}$ the total, the muon contribution is $\mathbf{j}_{\mathrm{med},\mu}=\mathbf{j}_{\rm med} \Pi_{T,\mu}/\Pi_T$. Hence a proton current induces a muon current $\mathbf{j}_{\mathrm{med},\mu}=-\mathbf{j}_p \Pi_{T,\mu}/\Pi_T$, which sources muonic vector bosons in proportion to $g_{V\mu}$. In other words, these bosons effectively couple to protons with $\hat{g}_{V\!p}=-g_{V\!\mu}\Pi_{T,\mu}/\Pi_T$.

As a result, vector emission is dominated by $np$ dipole radiation. Using the same soft-radiation estimate as for $\mu\mu$ bremsstrahlung, $d\mathcal{E}/d\omega\sim \hat{g}_{V\!p}^2 \Delta v^2\sim \hat{g}_{V\!p}^2 p_N^2/m_N^2$, neglecting the difference between proton and neutron momenta. With $\omega\sim T$, the radiated energy per collision is $\Delta \mathcal{E}\sim \hat{g}_{V\!p}^2 p_N^2 T/m_N^2$. The scattering rate in a degenerate Fermi liquid is $dN/dVdt\sim n_p n_n\sigma_{np} v_{\rm rel} (T/\mu_N)^3$, where the cross section $\sigma_{np}\sim 1/p_N^2$, the nucleon densities are $n_p\sim n_n\sim p_N^3$, and $v_{\rm rel}\sim p_N/m_N$. The factor $(T/\mu_N)^3$ accounts for degeneracy suppression for three particles, whereas the fourth is fixed by energy-momentum conservation. Since $\mu_N\sim p_N^2/m_N$, we estimate
\begin{equation}
    Q_V^{(pn)}\sim \Delta \mathcal{E} \frac{dN}{dVdt}\sim \hat{g}^2_{V\!p}T^4 p_N.
\end{equation}
Our companion paper~\cite{Fiorillo:2026long} derives the emission rate using Landau Fermi-liquid theory, substantiating this estimate. Our bounds are based on the full result.

For scalars, the same medium response induces an effective coupling to the electron charge density, $\mathcal{L}\supset - \hat{g}_{\phi e}\phi\,\bar e\gamma^0e$. An accelerated electron produces an electromagnetic perturbation that is rapidly screened by muons, whose coherent response is encoded in $\hat{g}_{\phi e}$. Crucially, the induced interaction scales with electron energy rather than mass and is therefore relativistically enhanced. In analogy with vectors, the energy loss rate in the $ee\to ee\phi$ channel scales as
\begin{equation}
    Q_{\phi}^{(ee)}\sim \frac{\alpha^2 \hat g_{\phi e}^{\,2}\,\mu_e^2\, T^4} {(1-\beta_e^2)^2\,m_M}\,.
\end{equation}
The $\gamma_e^4=(1-\beta_e^2)^{-2}$ enhancement combines the usual $\gamma_e^2$ growth of electromagnetic radiation with an additional 
$\gamma_e^2$ because longitudinal radiation lacks the $\sin^2\theta$ suppression, with $\theta\sim \gamma^{-1}$ \cite{Fiorillo:2026long}.

For completeness, we also consider pseudoscalars, for which muon bremsstrahlung dominates. Compared to the vector case, the amplitude contains an additional factor $g_{a\mu} k\, m_M/(\omega\mu_\mu^2)$, leading to $T^6$ scaling in the superconducting phase, and $T^{17/3}$ otherwise.

{\bf\textit{Muon depletion by long-range forces}}---For light mediators or large couplings, the potential sourced by dense muons is large enough to affect the muons themselves. For scalars, this feedback occurs only for parameter regions excluded by fifth-force tests~(see Fig.~\ref{fig:Bounds}), whereas for vectors such bounds are absent. When the force range is small compared with the NS radius, the vector potential is $U_\mu=(g_{V\!\mu}/m_V)^2 n_\mu$ (see Supplemental Material, SupM~\cite{supplementalmaterial}). Beta equilibrium implies $\mu_\mu=\mu_e$. For degenerate muons, the strength of the  backreaction is controlled by the dimensionless parameter $\eta_V=g_{V\!\mu}^2\mu_e^2/(3\pi^2m_V^2)$ (see SupM~\cite{supplementalmaterial}), with the condition $\eta_V=1$ defining the line labeled ``$\mu$ depletion'' in Fig.~\ref{fig:Bounds}. In the strongly depleted limit, $\mu_e-U_\mu\simeq m_\mu$, so $U_\mu\simeq\Delta_\mu\equiv\mu_e-m_\mu$ and the self-consistent density is $n_\mu\simeq\Delta_\mu m_V^2/g_{V\!\mu}^2$.

In this regime, the effective vector coupling to protons induced by the coherent muon response is suppressed as $\hat{g}_{V\!p}\sim g_{V\!\mu}\,n_{\mu}/{(n_{\mu}+n_e)}\sim \Delta_\mu m_V^2/(g_{V\!\mu} n_e)$. Incoherent processes then dominate and are largely insensitive to muon depletion. From Eq.~\eqref{eq:QV_mumu_generic}, adapted to a generic $\mu f$ scattering with $F_{\rm deg}\to T^3/\mu_\mu \mu_f m_M$ and $n_\mu^2 \to n_\mu n_f\sim n_\mu p_{{\rm F},f}^3$, we find
\begin{equation}
    Q_V^{(\mu f)}\sim \frac{g_{V\!\mu}^2 \alpha^2 p_{{\rm F},f}^3T^4}{m_M \mu_f}.
\end{equation}
In other words, the incoherent emissivity is not parametrically suppressed by muon depletion because weaker degeneracy suppression compensates for the reduced muon density. However, incoherent collisions alone provide bounds that are weaker by approximately an order of magnitude, explaining the step in the exclusion contour when it crosses the muon-depletion line in  Fig.~\ref{fig:Bounds}.

For sufficiently large coupling, muon depletion is so strong that the remaining population becomes nondegenerate. The transition occurs for $p_{{\rm F},\mu}^2/2m_\mu\lesssim T$ and is marked ``Nondegenerate $\mu$'' in Fig.~\ref{fig:Bounds}. The emissivity now scales as $g_{V\!\mu}^2 n_\mu$, without degeneracy suppression and since $n_\mu \propto g_{V\!\mu}^{-2}$, the rate is independent of the coupling. Thus, if the vector boson emissivity reaches the cooling constraint while the muons are still degenerate, the bound persists to arbitrarily large couplings.

For $m_V\lesssim 10^{-7}\,\mathrm{eV}$, however, the nondegenerate regime is reached \textit{before} $L_V$ becomes comparable with $L_\nu$. The resulting coupling-independent vector emissivity is then too small to compete with neutrino cooling, eliminating the bound.

\textbf{\textit{Other constraints}}---The cooling speed of the nascent NS in SN~1987A, inferred from the neutrino signal~\cite{Fiorillo:2023frv}, constrains all interaction structures~\cite{Caputo:2021rux,Blinov:2025aha}. The corresponding tree-level bounds for scalars and vectors are shown as solid green lines in Fig.~\ref{fig:Bounds}. For vectors lighter than $m_V\lesssim 2\times10^{-2}\,{\rm eV}$, the nondegenerate muons are again depleted, as detailed in SupM~\cite{supplementalmaterial}.

In contrast to vectors, spin-0 bosons can decay to two photons, here through a muon triangle loop. The resulting bounds are dashed in Fig.~\ref{fig:Bounds}. For $m_{a,\phi}\gtrsim 100\,\mathrm{keV}$, the dominant constraints come from SN~1987A and all SNe in the Universe~\cite{Caputo:2021rux}. Lighter bosons are constrained by globular clusters to $G_{\phi\gamma\gamma}<0.47\times 10^{-10}\,\mathrm{GeV}^{-1}$  \cite{Dolan:2022kul}. This translates to a bound on $g_{\phi\mu}$ shown in Table~\ref{tab:Bounds}, with an analogous bound for pseudoscalars.

Radiative decays after leaving the NS can contribute to the cosmic X-ray background (CXB) or produce an X-ray excess in the spectra of isolated NSs. However, we show in the SupM~\cite{supplementalmaterial} that these constraints are weaker than those from NS cooling.

On the other hand, a loop-induced photon coupling allows scalars to interact with charged particles. We show in the SupM~\cite{supplementalmaterial} that this induced coupling to electrons is suppressed by powers of $m_e/m_\mu$, whereas the proton coupling is $g_{\phi p}=\alpha^2 g_{\phi\mu}/4$, in agreement with Ref.~\cite{Garani:2021ysl}. Fifth-force limits \cite{Hoskins:1985tn, Smith:1999cr, Kapner:2006si, Geraci:2008hb, Sushkov:2011md, Yang:2012zzb, Chen:2014oda, Decca:2007yb, Blakemore:2021zna, Venugopalan:2024kgu, Chiu:2009fqu, Bezerra:2010pq, Banishev:2012kkb, Banishev:2014jka, Berge:2017ovy, Lee:2020zjt, Tan:2020vpf, Matias:2025vrq, Boynewicz:2025daj, Ren:2026hff} then apply to muonic scalars despite the absence of muons in ordinary matter, yielding the bounds shown in Fig.~\ref{fig:Bounds}. These limits surpass astrophysical ones for $m_\phi\lesssim 30\,\mathrm{meV}$. 

Similarly, muonic vectors acquire kinetic mixing with the photon through charged-fermion loops. With anomaly-free $L_\mu - L_\tau$ as our reference case, the opposite muon and tau charges cancel the ultraviolet divergence. For vanishing UV contributions, the infrared interaction is $\frac{\epsilon}{2}F^{\mu\nu}V_{\mu\nu}$ with $\epsilon=-[eg_{V\!\mu}/(6\pi^2)]\ln(m_\tau/m_\mu)\simeq-g_{V\!\mu}/70$. We use this relation to rescale astrophysical dark-photon bounds~\cite{Li:2023vpv,Dolan:2023cjs}. Other UV completions can modify $\epsilon$ through logarithmic threshold contributions from charged particles or an independent kinetic-mixing term, making these bounds model-dependent.

Fifth-force bounds do not apply to dark photons. Muonic vector forces can instead be probed through tests of Coulomb's law~\cite{Bartlett:1988yy, Kroff:2020zhp}, but these limits are weaker than astrophysical ones. 

Additional constraints arise from cosmological production through kinetic mixing. For $m_V\lesssim1\,{\rm MeV}$, the resulting irreducible freeze-in population
can decay into three photons, contributing to the diffuse and Galactic hard X-ray flux \cite{Cirelli:2010xx, Linden:2024fby, Caputo:2025avc}.

{\bf\textit{Discussion and outlook}}---As expected, NS cooling is a sensitive probe of muonic bosons. Surprisingly, however,
the large muon population emits scalars and vectors primarily through a coherent response to electron and proton collisions. The resulting bounds surpass previous sensitivity by many orders of magnitude over a broad range of parameters (Fig.~\ref{fig:Bounds}). Only for light scalars can loop-induced fifth forces provide stronger bounds. For sufficiently light vectors, on the other hand, muons are depleted by their own potential and NS cooling loses its sensitivity to the interaction.

This depletion may modify previous NS bounds \cite{Liu:2026dbd}, particularly when couplings to neutrinos are included. Such couplings were neglected here. Heavier bosons produced in SNe can decay to neutrinos en route to Earth, generating a high-energy flux constrained by the absence of corresponding events in the SN~1987A neutrino signal \cite{Fiorillo:2022cdq}. Previous studies of muonic bosons in SNe \cite{Croon:2020lrf, Caputo:2021rux, Akita:2023iwq, Fiorillo:2022cdq, Blinov:2025aha, Ferreira:2025qui, Das:2026uam} have neglected in-medium couplings, although our results suggest that these may significantly affect the emission. These intriguing questions point to several directions for future work.

{\bf\textit{Acknowledgments}}---This article is based upon work from COST Action COSMIC WISPers (CA21106), supported by COST (European Cooperation in Science and Technology). 
GGR acknowledges partial support by the German Research Foundation (DFG) through the Collaborative Research Centre ``Neutrinos and Dark Matter in Astro- and Particle Physics (NDM),'' Grant SFB--1258--283604770 and under Germany’s Excellence Strategy
through the Cluster of Excellence ORIGINS EXC--2094--390783311.
EV acknowledges support by the Italian MUR Departments of Excellence grant 2023-2027 ``Quantum Frontier''. 
AL, NS and EV are supported by the Italian MUR through the FIS 2 project FIS-2023-01577 (DD n. 23314 10-12-2024, CUP C53C24001460001), and by Istituto Nazionale di Fisica Nucleare (INFN) through the Theoretical Astroparticle Physics (TAsP) project.

\bibliographystyle{bibi}
\bibliography{references}

@article{Liu:2026dbd,
    author = "Liu, Yuxin and Liu, Zhen and Shkerin, Andrey and Shu, Jing and Zhao, Yue",
    title = "{New Gauge Forces, Neutron Stars and Schwinger Neutrino Production}",
    eprint = "2606.20393",
    archivePrefix = "arXiv",
    primaryClass = "hep-ph",
    month = "6",
    year = "2026"
}

@article{Tetzlaff:2012rz,
    author = "Tetzlaff, Nina and Schmidt, Janos G. and Hohle, Markus M. and Neuhaeuser, Ralph",
    title = "{Neutron stars from young nearby associations the origin of RXJ1605.3+3249}",
    eprint = "1202.1388",
    archivePrefix = "arXiv",
    primaryClass = "astro-ph.GA",
    doi = "10.1071/AS11057",
    journal = "Publ. Astron. Soc. Austral.",
    volume = "29",
    pages = "98",
    year = "2012"
}

@article{Pires:2019qsk,
    author = "Pires, A. M. and Schwope, A. D. and Haberl, F. and Zavlin, V. E. and Motch, C. and Zane, S.",
    title = "{A deep XMM-Newton look on the thermally emitting isolated neutron star RX J1605.3+3249}",
    eprint = "1901.08533",
    archivePrefix = "arXiv",
    primaryClass = "astro-ph.HE",
    doi = "10.1051/0004-6361/201834801",
    journal = "Astron. Astrophys.",
    volume = "623",
    pages = "A73",
    year = "2019"
}

@article{Krivonos:2020qvl,
    author = "Krivonos, Roman and Wik, Daniel and Grefenstette, Brian and Madsen, Kristin and Perez, Kerstin and Rossland, Steven and Sazonov, Sergey and Zoglauer, Andreas",
    title = "{$NuSTAR$ measurement of the cosmic X-ray background in the 3{\textendash}20 keV energy band}",
    eprint = "2011.11469",
    archivePrefix = "arXiv",
    primaryClass = "astro-ph.HE",
    doi = "10.1093/mnras/stab209",
    journal = "Mon. Not. Roy. Astron. Soc.",
    volume = "502",
    number = "3",
    pages = "3966--3975",
    year = "2021"
}

@article{Ren:2026hff,
    author = "Ren, Yi-Chong and Xu, Feng and Broer, Wijnand and Chen, Xiao-Jing and Xue, Fei",
    title = "{Field-Tunable Meissner-Levitated Ferromagnetic Microsphere Sensor for Cryogenic Casimir and Short-Range Gravity Tests}",
    eprint = "2602.13829",
    archivePrefix = "arXiv",
    primaryClass = "quant-ph",
    month = "2",
    year = "2026"
}

@article{Matias:2025vrq,
    author = "Matias, J. E. J. and Lemos, A. S. and Dahia, F.",
    title = "{Probing short-distance modifications of gravity via spin-independent and spin-dependent effects in muonic atoms}",
    eprint = "2511.00719",
    archivePrefix = "arXiv",
    primaryClass = "hep-ph",
    doi = "10.1140/epjp/s13360-026-07560-5",
    journal = "Eur. Phys. J. Plus",
    volume = "141",
    number = "3",
    pages = "328",
    year = "2026"
}

@article{Decca:2007yb,
    author = "Decca, R. S. and Lopez, D. and Fischbach, E. and Klimchitskaya, G. L. and Krause, D. E. and Mostepanenko, V. M.",
    title = "{Tests of new physics from precise measurements of the Casimir pressure between two gold-coated plates}",
    eprint = "hep-ph/0703290",
    archivePrefix = "arXiv",
    doi = "10.1103/PhysRevD.75.077101",
    journal = "Phys. Rev. D",
    volume = "75",
    pages = "077101",
    year = "2007"
}

@article{Boynewicz:2025daj,
    author = "Boynewicz, J. and Sackett, C. A.",
    title = "{Probing short-range gravity using quantum reflection}",
    eprint = "2511.08770",
    archivePrefix = "arXiv",
    primaryClass = "cond-mat.quant-gas",
    month = "11",
    year = "2025"
}

@article{Gruber:1999yr,
    author = "Gruber, D. E. and Matteson, J. L. and Peterson, L. E. and Jung, G. V.",
    title = "{The spectrum of diffuse cosmic hard x-rays measured with HEAO-1}",
    eprint = "astro-ph/9903492",
    archivePrefix = "arXiv",
    reportNumber = "SP-98-25",
    doi = "10.1086/307450",
    journal = "Astrophys. J.",
    volume = "520",
    pages = "124",
    year = "1999"
}

@article{Vitagliano:2019yzm,
    author = "Vitagliano, Edoardo and Tamborra, Irene and Raffelt, Georg",
    title = "{Grand Unified Neutrino Spectrum at Earth: Sources and Spectral Components}",
    eprint = "1910.11878",
    archivePrefix = "arXiv",
    primaryClass = "astro-ph.HE",
    reportNumber = "MPP-2019-205",
    doi = "10.1103/RevModPhys.92.045006",
    journal = "Rev. Mod. Phys.",
    volume = "92",
    pages = "45006",
    year = "2020"
}

@article{Dror:2019uea,
    author = "Dror, Jeff A. and Laha, Ranjan and Opferkuch, Toby",
    title = "{Probing muonic forces with neutron star binaries}",
    eprint = "1909.12845",
    archivePrefix = "arXiv",
    primaryClass = "hep-ph",
    reportNumber = "CERN-TH-2019-150",
    doi = "10.1103/PhysRevD.102.023005",
    journal = "Phys. Rev. D",
    volume = "102",
    number = "2",
    pages = "023005",
    year = "2020"
}

@book{Landau:1975pou,
    author = "Landau, L. D. and Lifschits, E. M.",
    title = "{The Classical Theory of Fields}",
    doi = "10.1016/c2009-0-14608-1",
    isbn = "978-0-08-018176-9",
    publisher = "Pergamon Press",
    address = "Oxford",
    series = "Course of Theoretical Physics",
    volume = "Volume 2",
    year = "1975"
}

@article{He:1991qd,
    author = "He, Xiao-Gang and Joshi, Girish C. and Lew, H. and Volkas, R. R.",
    title = "{Simplest $Z'$ model}",
    reportNumber = "CERN-TH-6084-91, UM-P-91-32, OZ-91-07",
    doi = "10.1103/PhysRevD.44.2118",
    journal = "Phys. Rev. D",
    volume = "44",
    pages = "2118--2132",
    year = "1991"
}

@article{He:1990pn,
    author = "He, X. G. and Joshi, Girish C. and Lew, H. and Volkas, R. R.",
    title = "{New-$Z'$ phenomenology}",
    reportNumber = "UM-P-90/42, OZ-P-90/16",
    doi = "10.1103/PhysRevD.43.R22",
    journal = "Phys. Rev. D",
    volume = "43",
    pages = "22--24",
    year = "1991"
}

@misc{XMMUHB_EPICSensitivity,
  author       = {{XMM-Newton Science Operations Centre}},
  title        = {{XMM-Newton Users Handbook: EPIC's sensitivity limits}},
  howpublished = {\url{https://xmm-tools.cosmos.esa.int/external/xmm_user_support/documentation/uhb/epicsens.html}},
  note         = {European Space Agency, XMM-Newton Users Handbook, Section 3.3.8; accessed 21 May 2026}
}

@article{Foot:1990mn,
    author = "Foot, Robert",
    title = "{New Physics From Electric Charge Quantization?}",
    reportNumber = "MAD/TH/90-14",
    doi = "10.1142/S0217732391000543",
    journal = "Mod. Phys. Lett. A",
    volume = "6",
    pages = "527--530",
    year = "1991"
}

@article{Foot:1994vd,
    author = "Foot, Robert and He, X. G. and Lew, H. and Volkas, R. R.",
    title = "{Model for a light $Z'$ boson}",
    eprint = "hep-ph/9401250",
    archivePrefix = "arXiv",
    reportNumber = "OITS-532, UM-P-93-115, OZ-93-26, IP-ASTP-32",
    doi = "10.1103/PhysRevD.50.4571",
    journal = "Phys. Rev. D",
    volume = "50",
    pages = "4571--4580",
    year = "1994"
}

@article{Blinov:2025aha,
    author = "Blinov, Nikita and Fox, Patrick J. and Kelly, Kevin J. and Plestid, Ryan and Zhou, Tao",
    title = "{L$_{\mu}$ {\ensuremath{-}} L$_{\tau}$ gauge bosons in beam dumps and supernovae}",
    eprint = "2511.09619",
    archivePrefix = "arXiv",
    primaryClass = "hep-ph",
    reportNumber = "FERMILAB-PUB-25-0818-T, MI-HET-869, CALT-TH/2025-034, CERN-TH-2025-229",
    doi = "10.1007/JHEP06(2026)155",
    journal = "JHEP",
    volume = "06",
    pages = "155",
    year = "2026"
}

@article{Akita:2023iwq,
    author = "Akita, Kensuke and Im, Sang Hui and Masud, Mehedi and Yun, Seokhoon",
    title = "{Limits on heavy neutral leptons, {$Z'$} bosons and majorons from high-energy supernova neutrinos}",
    eprint = "2312.13627",
    archivePrefix = "arXiv",
    primaryClass = "hep-ph",
    reportNumber = "CTPU-PTC-23-55",
    doi = "10.1007/JHEP07(2024)057",
    journal = "JHEP",
    volume = "07",
    pages = "057",
    year = "2024"
}

@article{Batell:2017kty,
    author = "Batell, Brian and Freitas, Ayres and Ismail, Ahmed and Mckeen, David",
    title = "{Flavor-specific scalar mediators}",
    eprint = "1712.10022",
    archivePrefix = "arXiv",
    primaryClass = "hep-ph",
    doi = "10.1103/PhysRevD.98.055026",
    journal = "Phys. Rev. D",
    volume = "98",
    number = "5",
    pages = "055026",
    year = "2018"
}

@article{Batell:2021xsi,
    author = "Batell, Brian and Freitas, Ayres and Ismail, Ahmed and McKeen, David and Rai, Mudit",
    title = "{Renormalizable models of flavor-specific scalars}",
    eprint = "2107.08059",
    archivePrefix = "arXiv",
    primaryClass = "hep-ph",
    reportNumber = "PITT-PACC-2114",
    doi = "10.1103/PhysRevD.104.115032",
    journal = "Phys. Rev. D",
    volume = "104",
    number = "11",
    pages = "115032",
    year = "2021"
}

@article{DiLuzio:2020wdo,
    author = "Di Luzio, Luca and Giannotti, Maurizio and Nardi, Enrico and Visinelli, Luca",
    title = "{The landscape of QCD axion models}",
    eprint = "2003.01100",
    archivePrefix = "arXiv",
    primaryClass = "hep-ph",
    reportNumber = "DESY 20-036, DESY-20-036",
    doi = "10.1016/j.physrep.2020.06.002",
    journal = "Phys. Rept.",
    volume = "870",
    pages = "1--117",
    year = "2020"
}

@article{KumarPoddar:2019ceq,
    author = "Kumar Poddar, Tanmay and Mohanty, Subhendra and Jana, Soumya",
    title = "{Vector gauge boson radiation from compact binary systems in a gauged $L_\mu-L_\tau$ scenario}",
    eprint = "1908.09732",
    archivePrefix = "arXiv",
    primaryClass = "hep-ph",
    doi = "10.1103/PhysRevD.100.123023",
    journal = "Phys. Rev. D",
    volume = "100",
    number = "12",
    pages = "123023",
    year = "2019"
}

@article{Croon:2020lrf,
    author = "Croon, Djuna and Elor, Gilly and Leane, Rebecca K. and McDermott, Samuel D.",
    title = "{Supernova Muons: New Constraints on $Z'$ Bosons, Axions and ALPs}",
    eprint = "2006.13942",
    archivePrefix = "arXiv",
    primaryClass = "hep-ph",
    reportNumber = "MIT-CTP/5214, FERMILAB-PUB-20-246-A-T",
    doi = "10.1007/JHEP01(2021)107",
    journal = "JHEP",
    volume = "01",
    pages = "107",
    year = "2021"
}

@article{Dolan:2022kul,
    author = "Dolan, Matthew J. and Hiskens, Frederick J. and Volkas, Raymond R.",
    title = "{Advancing globular cluster constraints on the axion-photon coupling}",
    eprint = "2207.03102",
    archivePrefix = "arXiv",
    primaryClass = "hep-ph",
    doi = "10.1088/1475-7516/2022/10/096",
    journal = "JCAP",
    volume = "10",
    pages = "096",
    year = "2022"
}

@article{Sushkov:2011md,
    author = "Sushkov, A. O. and Kim, W. J. and Dalvit, D. A. R. and Lamoreaux, S. K.",
    title = "{New Experimental Limits on Non-Newtonian Forces in the Micrometer Range}",
    eprint = "1108.2547",
    archivePrefix = "arXiv",
    primaryClass = "quant-ph",
    doi = "10.1103/PhysRevLett.107.171101",
    journal = "Phys. Rev. Lett.",
    volume = "107",
    pages = "171101",
    year = "2011"
}

@article{Iwamoto:1984ir,
    author = "Iwamoto, N.",
    title = "{Axion Emission from Neutron Stars}",
    doi = "10.1103/PhysRevLett.53.1198",
    journal = "Phys. Rev. Lett.",
    volume = "53",
    pages = "1198--1201",
    year = "1984"
}

@article{Buschmann:2021juv,
    author = "Buschmann, Malte and Dessert, Christopher and Foster, Joshua W. and Long, Andrew J. and Safdi, Benjamin R.",
    title = "{Upper Limit on the QCD Axion Mass from Isolated Neutron Star Cooling}",
    eprint = "2111.09892",
    archivePrefix = "arXiv",
    primaryClass = "hep-ph",
    doi = "10.1103/PhysRevLett.128.091102",
    journal = "Phys. Rev. Lett.",
    volume = "128",
    number = "9",
    pages = "091102",
    year = "2022"
}

@article{Blakemore:2021zna,
    author = "Blakemore, Charles P. and Fieguth, Alexander and Kawasaki, Akio and Priel, Nadav and Martin, Denzal and Rider, Alexander D. and Wang, Qidong and Gratta, Giorgio",
    title = "{Search for non-Newtonian interactions at micrometer scale with a levitated test mass}",
    eprint = "2102.06848",
    archivePrefix = "arXiv",
    primaryClass = "hep-ex",
    doi = "10.1103/PhysRevD.104.L061101",
    journal = "Phys. Rev. D",
    volume = "104",
    number = "6",
    pages = "L061101",
    year = "2021"
}

@article{Venugopalan:2024kgu,
    author = "Venugopalan, Gautam and others",
    title = "{Optomechanical vector sensing of new forces at 6 micron separation}",
    eprint = "2412.13167",
    archivePrefix = "arXiv",
    primaryClass = "hep-ex",
    doi = "10.1038/s41598-026-35656-6",
    journal = "Sci. Rep.",
    volume = "16",
    pages = "5180",
    year = "2026"
}

@article{Banishev:2014jka,
    author = "Banishev, A. A. and Wagner, J. and Emig, T. and Zandi, R. and Mohideen, U.",
    title = "{Experimental and theoretical investigation of the angular dependence of the Casimir force between sinusoidally corrugated surfaces}",
    eprint = "1402.2716",
    archivePrefix = "arXiv",
    primaryClass = "quant-ph",
    doi = "10.1103/PhysRevB.89.235436",
    journal = "Phys. Rev. B",
    volume = "89",
    number = "23",
    pages = "235436",
    year = "2014"
}

@article{Banishev:2012kkb,
    author = "Banishev, A. A. and Wagner, J. and Emig, T. and Zandi, R. and Mohideen, U.",
    title = "{Demonstration of Angle Dependent Casimir Force Between Corrugations}",
    eprint = "1212.6271",
    archivePrefix = "arXiv",
    primaryClass = "quant-ph",
    doi = "10.1103/PhysRevLett.110.250403",
    journal = "Phys. Rev. Lett.",
    volume = "110",
    number = "25",
    pages = "250403",
    year = "2013"
}

@article{Bezerra:2010pq,
    author = "Bezerra, V. B. and Klimchitskaya, G. L. and Mostepanenko, V. M. and Romero, C.",
    title = "{Advance and prospects in constraining the Yukawa-type corrections to Newtonian gravity from the Casimir effect}",
    eprint = "1002.2141",
    archivePrefix = "arXiv",
    primaryClass = "hep-th",
    doi = "10.1103/PhysRevD.81.055003",
    journal = "Phys. Rev. D",
    volume = "81",
    pages = "055003",
    year = "2010"
}

@ARTICLE{2001A&A...365L...1J,
       author = {{Jansen}, F. and {Lumb}, D. and {Altieri}, B. and {Clavel}, J. and {Ehle}, M. and {Erd}, C. and {Gabriel}, C. and {Guainazzi}, M. and {Gondoin}, P. and {Much}, R. and {Munoz}, R. and {Santos}, M. and {Schartel}, N. and {Texier}, D. and {Vacanti}, G.},
        title = "{XMM-Newton observatory. I. The spacecraft and operations}",
      journal = {\aap},
         year = 2001,
        month = jan,
       volume = {365},
        pages = {L1-L6},
          doi = {10.1051/0004-6361:20000036},
       adsurl = {https://ui.adsabs.harvard.edu/abs/2001A&A...365L...1J}
}

@article{Langhoff:2022bij,
    author = "Langhoff, Kevin and Outmezguine, Nadav Joseph and Rodd, Nicholas L.",
    title = "{Irreducible Axion Background}",
    eprint = "2209.06216",
    archivePrefix = "arXiv",
    primaryClass = "hep-ph",
    reportNumber = "CERN-TH-2022-148",
    doi = "10.1103/PhysRevLett.129.241101",
    journal = "Phys. Rev. Lett.",
    volume = "129",
    number = "24",
    pages = "241101",
    year = "2022"
}

@article{Chiu:2009fqu,
    author = "Chiu, H. -C. and Klimchitskaya, G. L. and Marachevsky, V. N. and Mostepanenko, V. M. and Mohideen, U.",
    title = "{Demonstration of the asymmetric lateral Casimir force between corrugated surfaces in the nonadditive regime}",
    eprint = "0909.2161",
    archivePrefix = "arXiv",
    primaryClass = "quant-ph",
    doi = "10.1103/PhysRevB.80.121402",
    journal = "Phys. Rev. B",
    volume = "80",
    number = "12",
    pages = "121402",
    year = "2009"
}

@article{Ishizuka:1989ts,
    author = "Ishizuka, N. and Yoshimura, M.",
    title = "{Axion and Dilaton Emissivity from Nascent Neutron Stars}",
    reportNumber = "TU-89-349",
    doi = "10.1143/PTP.84.233",
    journal = "Prog. Theor. Phys.",
    volume = "84",
    pages = "233--250",
    year = "1990"
}

@article{Fiorillo:2023frv,
    author = "Fiorillo, Damiano F. G. and Heinlein, Malte and Janka, Hans-Thomas and Raffelt, Georg and Vitagliano, Edoardo and Bollig, Robert",
    title = "{Supernova simulations confront SN 1987A neutrinos}",
    eprint = "2308.01403",
    archivePrefix = "arXiv",
    primaryClass = "astro-ph.HE",
    doi = "10.1103/PhysRevD.108.083040",
    journal = "Phys. Rev. D",
    volume = "108",
    number = "8",
    pages = "083040",
    year = "2023"
}

@article{Fiorillo:2025zzx,
    author = "Fiorillo, Damiano F. G. and Lella, Alessandro and O'Hare, Ciaran A. J. and Vitagliano, Edoardo",
    title = "{Leading Bounds on Micrometer to Picometer Fifth Forces from Neutron Star Cooling}",
    eprint = "2506.19906",
    archivePrefix = "arXiv",
    primaryClass = "hep-ph",
    reportNumber = "BARI-TH/776-25",
    doi = "10.1103/tlqz-713s",
    journal = "Phys. Rev. Lett.",
    volume = "135",
    number = "21",
    pages = "211003",
    year = "2025"
}

@article{Chen:2014oda,
    author = "Chen, Y. -J. and Tham, W. K. and Krause, D. E. and Lopez, D. and Fischbach, Ephraim and Decca, R. S.",
    title = "{Stronger Limits on Hypothetical Yukawa Interactions in the 30\textendash{}8000 nm Range}",
    eprint = "1410.7267",
    archivePrefix = "arXiv",
    primaryClass = "hep-ex",
    doi = "10.1103/PhysRevLett.116.221102",
    journal = "Phys. Rev. Lett.",
    volume = "116",
    number = "22",
    pages = "221102",
    year = "2016"
}

@article{Geraci:2008hb,
    author = "Geraci, Andrew A. and Smullin, Sylvia J. and Weld, David M. and Chiaverini, John and Kapitulnik, Aharon",
    title = "{Improved constraints on non-Newtonian forces at 10 microns}",
    eprint = "0802.2350",
    archivePrefix = "arXiv",
    primaryClass = "hep-ex",
    doi = "10.1103/PhysRevD.78.022002",
    journal = "Phys. Rev. D",
    volume = "78",
    pages = "022002",
    year = "2008"
}

@article{Kapner:2006si,
    author = "Kapner, D. J. and Cook, T. S. and Adelberger, E. G. and Gundlach, J. H. and Heckel, Blayne R. and Hoyle, C. D. and Swanson, H. E.",
    title = "{Tests of the gravitational inverse-square law below the dark-energy length scale}",
    eprint = "hep-ph/0611184",
    archivePrefix = "arXiv",
    doi = "10.1103/PhysRevLett.98.021101",
    journal = "Phys. Rev. Lett.",
    volume = "98",
    pages = "021101",
    year = "2007"
}

@article{Lee:2020zjt,
    author = "Lee, J. G. and Adelberger, E. G. and Cook, T. S. and Fleischer, S. M. and Heckel, B. R.",
    title = "{New Test of the Gravitational $1/r^2$ Law at Separations down to 52 $\mu$m}",
    eprint = "2002.11761",
    archivePrefix = "arXiv",
    primaryClass = "hep-ex",
    doi = "10.1103/PhysRevLett.124.101101",
    journal = "Phys. Rev. Lett.",
    volume = "124",
    number = "10",
    pages = "101101",
    year = "2020"
}

@article{Smith:1999cr,
    author = "Smith, G. L. and Hoyle, C. D. and Gundlach, J. H. and Adelberger, E. G. and Heckel, Blayne R. and Swanson, H. E.",
    title = "{Short range tests of the equivalence principle}",
    eprint = "2405.10982",
    archivePrefix = "arXiv",
    primaryClass = "gr-qc",
    doi = "10.1103/PhysRevD.61.022001",
    journal = "Phys. Rev. D",
    volume = "61",
    pages = "022001",
    year = "2000"
}

@article{Yang:2012zzb,
    author = "Yang, Shan-Qing and Zhan, Bi-Fu and Wang, Qing-Lan and Shao, Cheng-Gang and Tu, Liang-Cheng and Tan, Wen-Hai and Luo, Jun",
    title = "{Test of the Gravitational Inverse Square Law at Millimeter Ranges}",
    doi = "10.1103/PhysRevLett.108.081101",
    journal = "Phys. Rev. Lett.",
    volume = "108",
    pages = "081101",
    year = "2012"
}

@article{Tan:2020vpf,
    author = "Tan, Wen-Hai and others",
    title = "{Improvement for Testing the Gravitational Inverse-Square Law at the Submillimeter Range}",
    doi = "10.1103/PhysRevLett.124.051301",
    journal = "Phys. Rev. Lett.",
    volume = "124",
    number = "5",
    pages = "051301",
    year = "2020"
}

@article{Hoskins:1985tn,
    author = "Hoskins, J. K. and Newman, R. D. and Spero, R. and Schultz, J.",
    title = "{Experimental tests of the gravitational inverse square law for mass separations from 2-cm to 105-cm}",
    doi = "10.1103/PhysRevD.32.3084",
    journal = "Phys. Rev. D",
    volume = "32",
    pages = "3084--3095",
    year = "1985"
}

@article{Berge:2017ovy,
    author = "Berg\'e, Joel and Brax, Philippe and M\'etris, Gilles and Pernot-Borr\`as, Martin and Touboul, Pierre and Uzan, Jean-Philippe",
    title = "{MICROSCOPE Mission: First Constraints on the Violation of the Weak Equivalence Principle by a Light Scalar Dilaton}",
    eprint = "1712.00483",
    archivePrefix = "arXiv",
    primaryClass = "gr-qc",
    doi = "10.1103/PhysRevLett.120.141101",
    journal = "Phys. Rev. Lett.",
    volume = "120",
    number = "14",
    pages = "141101",
    year = "2018"
}

@article{Linden:2024fby,
    author = "Linden, Tim and Nguyen, Thong T. Q. and Tait, Tim M. P.",
    title = "{X-ray constraints on dark photon tridents}",
    eprint = "2406.19445",
    archivePrefix = "arXiv",
    primaryClass = "hep-ph",
    reportNumber = "UCI-HEP-TR-2024-10",
    doi = "10.1103/37gn-x3y1",
    journal = "Phys. Rev. D",
    volume = "112",
    number = "2",
    pages = "023026",
    year = "2025"
}

@article{Cirelli:2010xx,
    author = "Cirelli, Marco and Corcella, Gennaro and Hektor, Andi and Hutsi, Gert and Kadastik, Mario and Panci, Paolo and Raidal, Martti and Sala, Filippo and Strumia, Alessandro",
    title = "{PPPC 4 DM ID: A Poor Particle Physicist Cookbook for Dark Matter Indirect Detection}",
    eprint = "1012.4515",
    archivePrefix = "arXiv",
    primaryClass = "hep-ph",
    reportNumber = "CERN-PH-TH-2010-057, SACLAY-T10-025, IFUP-TH-2010-44",
    doi = "10.1088/1475-7516/2012/10/E01",
    journal = "JCAP",
    volume = "03",
    pages = "051",
    year = "2011",
    note = "[Erratum: JCAP 10, E01 (2012)]"
}

@article{Caputo:2025avc,
    author = "Caputo, Andrea and Park, Jaeyoung and Yun, Seokhoon",
    title = "{Heavy dark photon handbook: Cosmological and astrophysical bounds}",
    eprint = "2511.15785",
    archivePrefix = "arXiv",
    primaryClass = "hep-ph",
    reportNumber = "CTPU-PTC-25-42",
    doi = "10.1103/m6zs-jxxp",
    journal = "Phys. Rev. D",
    volume = "113",
    number = "7",
    pages = "075001",
    year = "2026"
}

@article{Garani:2021ysl,
    author = "Garani, Raghuveer and Gasparotto, Federico and Mastrolia, Pierpaolo and Munch, Henrik J. and Palomares-Ruiz, Sergio and Primo, Amedeo",
    title = "{Two-photon exchange in leptophilic dark matter scenarios}",
    eprint = "2105.12116",
    archivePrefix = "arXiv",
    primaryClass = "hep-ph",
    reportNumber = "ZU-TH 21/21",
    doi = "10.1007/JHEP12(2021)212",
    journal = "JHEP",
    volume = "12",
    pages = "212",
    year = "2021"
}

@article{Ferreira:2025qui,
    author = "Ferreira, Ricardo Z. and Marsh, M. C. David and Ravensburg, Eike",
    title = "{ALP couplings to muons and electrons: A comprehensive analysis of supernova bounds}",
    eprint = "2510.14469",
    archivePrefix = "arXiv",
    primaryClass = "hep-ph",
    doi = "10.1016/j.dark.2026.102435",
    journal = "Phys. Dark Univ.",
    volume = "53",
    pages = "102435",
    year = "2026"
}

@article{Das:2026uam,
    author = "Das, Debottam and Ghosh, Purusottam and Puri, Rahul",
    title = "{SN1987A Constraints of Light $\boldsymbol{Z'}$ with Non-Mixing Polarisations}",
    eprint = "2606.23769",
    archivePrefix = "arXiv",
    primaryClass = "hep-ph",
    month = "6",
    year = "2026"
}

@misc{2016ascl.soft09009P,
       author = {{Page}, Dany},
        title = "{NSCool: Neutron star cooling code}",
 howpublished = {Astrophysics Source Code Library, record ascl:1609.009},
         year = 2016,
        month = sep,
          eid = {ascl:1609.009},
       adsurl = {https://ui.adsabs.harvard.edu/abs/2016ascl.soft09009P}
}

@article{Friman:1979ecl,
    author = "Friman, B. L. and Maxwell, O. V.",
    title = "{Neutron Star Neutrino Emissivities}",
    doi = "10.1086/157313",
    journal = "Astrophys. J.",
    volume = "232",
    pages = "541--557",
    year = "1979"
}

@article{Bottaro:2024ugp,
    author = "Bottaro, Salvatore and Caputo, Andrea and Fiorillo, Damiano F. G.",
    title = "{Neutrino emission in cold neutron stars: Bremsstrahlung and modified urca rates reexamined}",
    eprint = "2406.18640",
    archivePrefix = "arXiv",
    primaryClass = "hep-ph",
    reportNumber = "CERN-TH-2024-092",
    doi = "10.1088/1475-7516/2024/11/015",
    journal = "JCAP",
    volume = "11",
    pages = "015",
    year = "2024"
}

@article{Dolan:2023cjs,
    author = "Dolan, Matthew J. and Hiskens, Frederick J. and Volkas, Raymond R.",
    title = "{Constraining dark photons with self-consistent simulations of globular cluster stars}",
    eprint = "2306.13335",
    archivePrefix = "arXiv",
    primaryClass = "hep-ph",
    doi = "10.1088/1475-7516/2024/05/099",
    journal = "JCAP",
    volume = "05",
    pages = "099",
    year = "2024"
}

@article{Li:2023vpv,
    author = "Li, Shao-Ping and Xu, Xun-Jie",
    title = "{Production rates of dark photons and Z' in the Sun and stellar cooling bounds}",
    eprint = "2304.12907",
    archivePrefix = "arXiv",
    primaryClass = "hep-ph",
    doi = "10.1088/1475-7516/2023/09/009",
    journal = "JCAP",
    volume = "09",
    pages = "009",
    year = "2023"
}

@article{Fiorillo:2026long,
    author = "Fiorillo, Damiano F. G. and Lella, Alessandro and Raffelt, Georg G. and Selimovi{\'c}, Nud{\v z}eim and Vitagliano, Edoardo",
    title = "{Production of Leptophilic Bosons in Ultradegenerate Relativistic Matter}",
    eprint = "2605.24081",
    archivePrefix = "arXiv",
    primaryClass = "hep-ph",
    month = "5",
    year = "2026"
}

@article{Fiorillo:2022cdq,
    author = "Fiorillo, Damiano F. G. and Raffelt, Georg G. and Vitagliano, Edoardo",
    title = "{Strong Supernova 1987A Constraints on Bosons Decaying to Neutrinos}",
    eprint = "2209.11773",
    archivePrefix = "arXiv",
    primaryClass = "hep-ph",
    doi = "10.1103/PhysRevLett.131.021001",
    journal = "Phys. Rev. Lett.",
    volume = "131",
    number = "2",
    pages = "021001",
    year = "2023"
}

@article{Bollig:2020xdr,
    author = "Bollig, Robert and DeRocco, William and Graham, Peter W. and Janka, Hans-Thomas",
    title = "{Muons in Supernovae: Implications for the Axion-Muon Coupling}",
    eprint = "2005.07141",
    archivePrefix = "arXiv",
    primaryClass = "hep-ph",
    doi = "10.1103/PhysRevLett.125.051104",
    journal = "Phys. Rev. Lett.",
    volume = "125",
    number = "5",
    pages = "051104",
    year = "2020",
    note = "Erratum \href{https://doi.org/10.1103/PhysRevLett.126.189901}{{\em Phys. Rev. Lett.} {\bf 126}, 189901 (2021)}"
}

@article{Caputo:2021rux,
    author = "Caputo, Andrea and Raffelt, Georg and Vitagliano, Edoardo",
    title = "{Muonic boson limits: Supernova redux}",
    eprint = "2109.03244",
    archivePrefix = "arXiv",
    primaryClass = "hep-ph",
    reportNumber = "MPP-2021-154",
    doi = "10.1103/PhysRevD.105.035022",
    journal = "Phys. Rev. D",
    volume = "105",
    number = "3",
    pages = "035022",
    year = "2022"
}

@article{Akmal:1998cf,
    author = "Akmal, A. and Pandharipande, V. R. and Ravenhall, D. G.",
    title = "{The Equation of state of nucleon matter and neutron star structure}",
    eprint = "nucl-th/9804027",
    archivePrefix = "arXiv",
    doi = "10.1103/PhysRevC.58.1804",
    journal = "Phys. Rev. C",
    volume = "58",
    pages = "1804--1828",
    year = "1998"
}

@article{Bartlett:1988yy,
    author = "Bartlett, D. F. and Loegl, S.",
    title = "{Limits on an Electromagnetic Fifth Force}",
    doi = "10.1103/PhysRevLett.61.2285",
    journal = "Phys. Rev. Lett.",
    volume = "61",
    pages = "2285--2287",
    year = "1988"
}

@article{Kroff:2020zhp,
    author = "Kroff, D. and Malta, P. C.",
    title = "{Constraining hidden photons via atomic force microscope measurements and the Plimpton-Lawton experiment}",
    eprint = "2008.02209",
    archivePrefix = "arXiv",
    primaryClass = "hep-ph",
    doi = "10.1103/PhysRevD.102.095015",
    journal = "Phys. Rev. D",
    volume = "102",
    number = "9",
    pages = "095015",
    year = "2020"
}

@misc{supplementalmaterial,
  note = {See Supplemental Material for a discussion of possible constraints from scalar decays in photon pairs, details on our calculation for the loop-induced scalar coupling to protons and an analysis of backraction effects induced by a strong vector-mediated muonic potential.},
  howpublished = {\url{https://}} 
}

\clearpage

\onecolumngrid

\bigskip

\begin{center}
\textbf{\large End Matter}
\end{center}

\smallskip

\twocolumngrid

{\bf\textit{Reference neutron star model}}---To illustrate the properties of medium-aged NSs used to constrain particle emission, we here show the properties of the NS model that best fits the observed age and luminosity of the pulsar J1605. The cooling lightcurves are calculated with the public code {\tt NSCool} \cite{2016ascl.soft09009P} by assuming the standard APR EoS without superfluidity, a NS mass $M_{\rm NS}=1.8\,M_\odot$ and a fraction $\Delta M/M=10^{-14}$ of light elements accreted onto the NS envelope. The NS profile refers to the epoch when $L_\nu=L_\gamma$ which occurs at $t\simeq7\times10^{4}$ years. The mass coordinate represents enclosed baryonic mass. All parameters are local quantities and as such relevant for the calculation of the emission rates. The core is very nearly isothermal, corresponding to variation of the local $T$ according to the gravitational potential. In the main text, luminosities are those for a distant observer. 

\onecolumngrid

\begin{figure}[ht]
\vskip8pt
  \centering
  \includegraphics[angle=-90, width=0.97\textwidth]{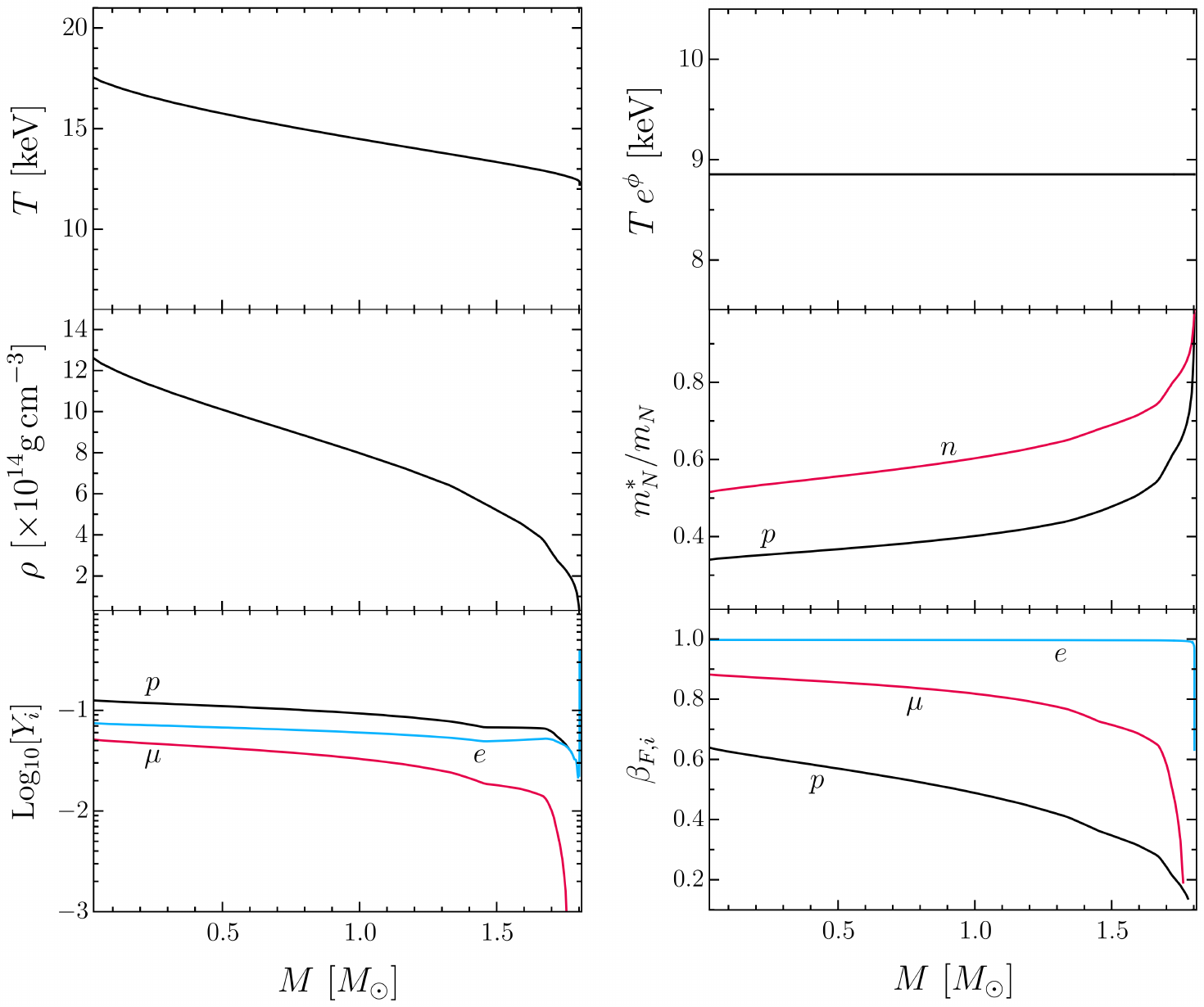}
  \caption{Profile of our reference NS model  at the epoch when $L_\nu=L_\gamma$,  shown as a function of enclosed baryonic mass. The upper panels refer to the local temperature and the redshifted temperature as measured by a distant observer. The middle panels show the NS density and nucleon effective masses, while the lower panels report particle fractions and Fermi velocities of the charged species present in the stellar plasma.\label{fig:Profiles}}
  \vskip-2pt
\end{figure}

\twocolumngrid

For simple estimates, we sometimes need to use typical average conditions that are roughly compatible with this model. We use an average density of $8.5\times10^{14}~{\rm g}~{\rm cm}^{-3}$ and $T=8.9~{\rm keV}$. Taking a one-zone proton abundance of $Y_p=0.110$ and imposing charge neutrality and beta equilibrium gives an electron chemical potential of $\mu_e=200~{\rm MeV}$, $Y_e=0.068$ and $Y_\mu=0.042$, and a relative muon abundance of $y=Y_\mu/Y_e=0.61$.

If muons are fully depleted by a repulsive long-range force while the baryon density and proton abundance are held fixed, charge neutrality implies $Y_e=Y_p=0.110$ and therefore $\mu_e=234~{\rm MeV}$.

\onecolumngrid

\clearpage

\onecolumngrid
\appendix

\setcounter{equation}{0}
\setcounter{figure}{0}
\setcounter{table}{0}
\setcounter{page}{1}
\makeatletter
\renewcommand{\theequation}{S\arabic{equation}}
\renewcommand{\thefigure}{S\arabic{figure}}
\renewcommand{\thepage}{S\arabic{page}}

\begin{center}
\textbf{\large Supplemental Material for the Letter\\[0.5ex]
{\em Neutron Star Bounds on Muonic Fifth Forces from Picometer to Meter Scales}}
\end{center}

\bigskip

In this Supplemental Material, we estimate possible constraints from scalar boson decays into photon pairs, detail our calculation for the loop-induced scalar coupling to protons and electrons and illustrate backreaction effects on the muon abundance induced by a strong vector-mediated potential.

\bigskip

\twocolumngrid

\section{A. Contribution to CXB}
(Pseudo)scalars will decay to photon pairs after leaving the NS. To estimate their contribution to the cosmic X-ray background (CXB), we consider bosons with $L_\phi\propto T^p$, which are assumed to begin dominating over neutrinos at $t=t_0$, when $L_\gamma=L_\nu=L_0$. At earlier times, they are only a perturbative effect. During neutrino cooling, $T\propto t^{-1/6}$, and therefore $L_\phi=L_0 (t/t_0)^{-p/6}$. The total energy emitted by bosons is $U_\phi=6 L_0 t_0/(6-p)$.

We assume a Maxwell-Boltzmann spectrum at the NS internal temperature $T$, so the average $\phi$ energy is $3T$ and the number emission rate $\dot N_\phi=L_\phi/3T$. At photon-neutrino equality, $T=T_0$, and $T=T_0 (t/t_0)^{-1/6}$ during neutrino cooling, so that $\dot N_\phi=(L_0/3T_0)(t/t_0)^{-(p-1)/6}$. Therefore, the total number of particles emitted is $N_\phi=[2/(7-p)]\,L_0 t_0/T_0$ and the average particle energy is $\langle E_\phi\rangle = U_\phi/N_\phi=3(7-p)T_0/(6-p)$. We approximate the integrated spectrum also as Maxwell-Boltzmann with $T_\phi=\langle E_\phi\rangle/3=(7-p)T_0/(6-p)$.

After neutrino-photon equality, the NS cools mainly by photon emission and we conservatively neglect bosons emitted later, which also have smaller energies. As reference parameters we use $L_0=2\times10^{33}~{\rm erg}~{\rm s}^{-1}$, $T_0=10^8~{\rm K}=8.6~{\rm keV}$, and $t_0=0.7\times10^5$~years, implying $U_\phi=1.65\times 10^{55}~{\rm keV}/(6-p)$, $N_\phi=6.4\times 10^{53}/(7-p)$, and $\langle E_\phi \rangle=3T_\phi=25.8\,\mathrm{keV}\,(7-p)/(6-p)$.

The cosmic comoving density of all past core-collapse SNe was estimated in Eq.~(40) of Ref.~\cite{Vitagliano:2019yzm} as $n_{\rm cc}=\xi_{\rm cc}\,10^7~{\rm Mpc}^{-3}$, where $\xi_{\rm cc}=0.58$--1.05, depending on the assumed star-formation rate. Ignoring BH-forming SNe, the cosmic average NS density is roughly identical with $n_{\rm cc}$. The star formation rate was much larger in the past, so most NSs are very old. With $\xi_{\rm cc}=0.58$, the average cosmic density of the new particles is
\begin{equation}
  n_\phi\simeq n_{\rm cc} N_\phi\simeq 1.3\times10^{-13}~{\rm cm}^{-3}/(7-p).
\end{equation}
After multiplying with the speed of light, this is an isotropic flux of $3.8\times10^{-3}~{\rm cm}^{-2}~{\rm s}^{-1}/(7-p)$. To obtain the directional flux, one needs to divide by $4\pi$ (see Appendix~A of Ref.~\cite{Vitagliano:2019yzm}) so that finally $d\Phi_\phi/d\Omega\simeq3.0\times10^{-4}~{\rm cm}^{-2}~{\rm s}^{-1}~{\rm sr}^{-1}/(7-p)$.

Assuming conservatively that all of the new particles decay into two photons until the present epoch, the $\gamma$ spectrum normalized to~2 is
\begin{equation}
  \frac{df_\gamma}{dE_\gamma}=\int_{E_\gamma}^{\infty}dE_\phi\,\frac{2}{E_\phi}\,\frac{df_\phi}{dE_\phi}=
  \frac{E_\gamma+T_\phi}{T_\phi^2}\,e^{-E_\gamma/T_\phi},
\end{equation}
assuming the original particles had a Maxwell-Boltzmann spectrum at temperature $T_\phi$ that was normalized to~1. The average photon energy is $\langle E_\gamma\rangle=3T_\phi/2$ as expected. Most core-collapse SNe occur at cosmic redshift $z\simeq 1$, so for the normalized-to-2 present-day spectrum, we should use $T_\phi=T_{\phi,{\rm emission}}/(1+z)$, in our case $T_\phi=\frac{1}{1+z}\times\frac{1}{3}\times25.8\,\mathrm{keV}\,(7-p)/(6-p)\simeq 4.3\,\mathrm{keV}\,(7-p)/(6-p)$, implying $\langle E_\gamma\rangle\simeq 6.5~{\rm keV}~(7-p)/(6-p)$.

In terms of the present-day $T_\phi$, the energy spectrum instead of the number spectrum is
\begin{equation}\label{eq:theoryflux}
  S(E_\gamma)= \frac{d\Phi_\phi}{d\Omega}\,\frac{E_\gamma(E_\gamma+T_\phi)}{T_\phi^2}\,e^{-E_\gamma/T_\phi}
\end{equation}
which has units ${\rm keV}~{\rm keV}^{-1}~{\rm cm}^{-2}~{\rm s}^{-1}~{\rm sr}^{-1}$. 
The CXB has been measured by several instruments. In the 3--60~keV range, an analytic formula was given by Gruber et al.\ \cite{Gruber:1999yr},
\begin{equation}
  S(E_\gamma)=7.877\,E_{\rm keV}^{-0.29} e^{-E_{\rm keV}/41.13}\,
  \frac{{\rm keV}}{{\rm keV}~{\rm cm}^{2}~{\rm s}~{\rm sr}},
\end{equation}
which seems to agree with the more recent NuSTAR measurements \cite{Krivonos:2020qvl}.

For $p=4$, corresponding to a scalar boson, the maximum of the $\gamma$ flux from particle decays of Eq.~\eqref{eq:theoryflux} is around $1\times10^{-4}$ smaller than the measurements at $E_\gamma\simeq15$~keV, where the curves come closest. The situation is similar for $p=4-1/3=11/3$ that pertains to emission in the normal-conducting phase.

For $p=6$, corresponding to a pseudoscalar, the total energy $U_\phi$ formally diverges, due to a logarithmic contribution from early times. This points to the predominance of pseudoscalars emitted from the SN phase and one should not expect decay bounds from cold NSs to surpass those from SNe. Imposing a cut at an early time, where the temperature was a few hundred keV, modifies the above estimates only by a logarithmic factor of small relevance.

These generic estimates show that the CXB cannot constrain (pseudo)scalars beyond the NS cooling bounds.

\section{B. Decays from isolated neutron stars}
Scalars produced in nearby isolated NSs partly decay on their way to Earth, giving rise to an additional X-ray flux. To estimate this contribution, we take as an example our reference NS J1605 at a distance $D\simeq350\,{\rm pc}$ and age $t\simeq4.4\times10^{5}\,$ years~\cite{Tetzlaff:2012rz}, that would have an internal $T\simeq 4\,{\rm keV}$ in the frame of a distant observer.

The two-photon decay length of a scalar particle in the laboratory frame is
%%%%%%%%%%%%%%%%%%%%%%%%%%%%%%%%%%%%%
\begin{equation}
    \lambda_{\phi\gamma\gamma}=\frac{64\pi E_\phi}{g_{\phi\gamma\gamma}^2 m_{\phi}^4}
    =\frac{144\pi^3 m_\mu^2 E_\phi}{g_{\phi\mu}^2 \alpha^2 m_\phi^4}
    \,,
\end{equation}
%%%%%%%%%%%%%%%%%%%%%%%%%%%%%%%%%%%%%
where $g_{\phi\gamma\gamma}=2\alpha g_{\phi\mu}/3\pi m_\mu$ is the loop-induced two-photon coupling of muon-philic scalars~\cite{Caputo:2021rux}. Numerically, at the cooling bound $g_{\phi\mu}=10^{-12}$ and for $E_\phi=m_\phi=10~{\rm keV}$, we find $\lambda_{\phi\gamma\gamma}= 6\times10^{12}~{\rm pc}$, much larger than the distance.

After the epoch of photon-neutrino equality, the temperature in the core decays exponentially with time $T\simeq T_0(t/t_0)^{-1/6}\exp[-t/\tau]$ in which $\tau\simeq7\times10^{5}$, leading to $L_\phi\simeq L_0(t/t_0)^{-2/3}\exp(-4t/\tau) \exp(-m_\phi/T)$. In this expression, we accounted for an additional exponential suppression factor for the emission of bosons with masses $m_\phi\gtrsim T$. The rate of emitted particles at present is then estimated to $\dot{N}_\phi\simeq L_\phi/(m_\phi+3T/2)\simeq5.5\times10^{39}\,{\rm s}^{-1}$. 

A very simple argument showing that decay bounds are rather weak is based on energetics. For a coupling equal to the cooling bound, the present-day scalar luminosity from the NS is roughly $L_\gamma\simeq3\times10^{31}\,\mathrm{erg/s}$. Therefore, the energy flux at Earth in the keV energy window ($E_\phi\sim 1\,\mathrm{keV}$) is roughly $\Phi_\phi\sim L_\gamma/4\pi D^2 E_\phi\sim 10^{-12}\,\mathrm{erg/s\,cm}^2\,\mathrm{keV}$. A fraction $D/\lambda_{\phi\gamma\gamma}$ of this energy may have converted into visible X-rays, but as we have seen, this fraction is, even at a boson mass of $m_\phi=10\,\mathrm{keV}$, rather small, of the order of $10^{-10}$. For comparison, the sensitivity of XMM-Newton~\cite{2001A&A...365L...1J} reaches at most $10^{-16}\,\mathrm{erg/s\,cm}^2\,\mathrm{keV}$~\cite{XMMUHB_EPICSensitivity}. Hence, decays from isolated NSs are generally not a competitive probe compared with cooling.
\\
\\
\begin{figure}[h]
  \centering
  \includegraphics[width=0.8\columnwidth]{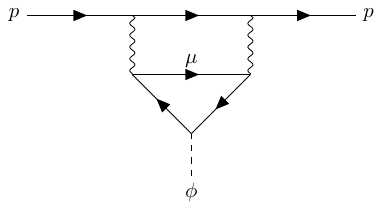}
  \caption{Two-loop muonic scalar interaction with protons.}\label{fig:muonloop}
\end{figure}

\section{C.~Loop-induced couplings to protons}
We consider the effective zero-momentum coupling of nonrelativistic nucleons to a scalar muonic boson. The amplitude, shown in Fig.~\ref{fig:muonloop}, can be written~as
\begin{widetext}
\begin{equation}
    -i F(K)=\lim_{K\to 0}\int \frac{d^4 Q}{(2\pi)^4}[-i \mathcal{T}^{\mu\nu}(K, Q)]\frac{16\pi^2 \alpha}{Q^2 (Q+K)^2}[\gamma^\mu i G(P-Q)\gamma^\nu + \gamma^\nu i G(P+Q+K)\gamma^\mu].
\end{equation}
\end{widetext}
Here $G(P)=1/(\slashed{P}-m_N)$ is the nucleon propagator, and $\mathcal{T}^{\mu\nu}(K, Q)$ denotes the muonic triangle loop amplitude. In the non-relativistic limit, we may only take the charge operator, so that $\mu=\nu=0$, and replace the propagator by its non-relativistic form. Taking the limit $K\to 0$, the two propagators add to $-2\pi i \delta(Q^0)$---a static field cannot carry energy. Therefore, after integrating the delta function, this takes the much simpler form
\begin{equation}
    -iF(K=0)=-i\int \frac{d^3\bq}{(2\pi)^3}\frac{16\pi^2\alpha}{\bq^4}\mathcal{T}^{00}(0,Q)\,.
\end{equation}
We now discuss the triangle loop, whose general form is
\begin{widetext}
\begin{eqnarray}\label{eq:loop_function}
    -i\mathcal{T}^{\mu\nu}(0,Q)=-\alpha g_{\phi\mu}\int \frac{d^4L}{(2\pi)^4}\,\frac{\mathrm{Tr}[\gamma^\mu(\slashed{L}+m_\mu)(\slashed{L}+m_\mu)\gamma^\nu (\slashed{L}-\slashed{Q}+m_\mu)]}{(L^2-m_\mu^2)^2[(L-Q)^2-m_\mu^2]} .
\end{eqnarray}
The integral must satisfy Ward's identity, and therefore must take the form
\begin{equation}
    \mathcal{T}^{\mu\nu}=T(Q^2)(Q^\mu Q^\nu-Q^2 g^{\mu\nu}).
\end{equation}
By taking the trace of Eq.~\eqref{eq:loop_function}, we then obtain
\begin{equation}
    3i Q^2 T(Q^2)=-\alpha g_{\phi\mu}[\Phi(Q)-\Phi(0)]
\end{equation}
with
\begin{equation}
    \Phi(Q)=16m_\mu \int \frac{d^4 L}{(2\pi)^4}\frac{m_\mu^2+L\cdot Q}{(L^2-m_\mu^2)^2[(L-Q)^2-m_\mu^2]}.
\end{equation}
Notice that the term $-\Phi(0)$ was originally absent from our loop, and has been introduced to renormalize the result and remove the unphysical contact term for $Q\to 0$. The function $\Phi(Q)$ can now be deduced  through the Feynman parameterization as
\begin{equation}
    \Phi(Q)=-\frac{im_\mu}{\pi^2}\int_0^1 dx x \frac{m_\mu^2+(1-x)Q^2}{m_\mu^2-x(1-x)Q^2}\,.
\end{equation}

Therefore, we finally have
\begin{equation}
    T(Q^2)=\frac{\alpha g_{\phi\mu}m_\mu}{2\pi^2}\int_0^1 dx \frac{x(1-x)}{m_\mu^2-x(1-x)Q^2},
\end{equation}
or explicitly performing the integral in Eq.~\eqref{eq:loop_function}, one finds
\begin{equation}
    T(Q^2) = g_{\phi\mu}\frac{\alpha}{8\pi}\frac{16m_\mu}{(Q^2-4m_\mu^2)Q^2}\left[\!4m_\mu^2-Q^2 + \frac{2m_\mu^2\sqrt{Q^2(Q^2-4m_\mu^2)}}{Q^2}\log\left(\!\frac{2m_\mu^2-Q^2+\sqrt{Q^2(Q^2-4m_\mu^2)}}{2m_\mu^2}\!\right)\!\right].
    \label{Eq:T(Q2)_ns}
\end{equation}
Notice that this has a finite limit for $Q\to 0$.
After evaluating this for $Q^0=0$, we finally obtain the effective coupling $F(K=0)$ as
\begin{align}
    g_{\phi p}=F(K=0) 
    =8\alpha^2 g_{\phi\mu}m_\mu \int \frac{d^3\bq}{(2\pi)^3\bq^2}\int_0^1 dx \frac{x(1-x)}{m_\mu^2+x(1-x)\bq^2}
    =\frac{1}{4}g_{\phi\mu}\alpha^2\,.
\end{align}
\end{widetext}

\section{D.~Loop-induced couplings to electrons}
Let us also consider the effective coupling induced to electrons. In this case, we must abandon the nonrelativistic approximation for the electron propagator; even if the interacting electrons are nonrelativistic, the momentum $Q$ flowing through them can reach up to the mass of the muon, much larger than that of the electron. Thus, we must use the full four-dimensional structure of the tensor $\mathcal{T}^{\mu\nu}$, but we can still take the limit $K\to 0$ and then obtain
\begin{widetext}
\begin{equation}
    F(K=0)=i\int \frac{d^4Q}{(2\pi)^4}\frac{16\pi^2\alpha T(Q^2)(Q_\mu Q_\nu-g_{\mu\nu} Q^2)}{Q^4}\left[\frac{\gamma^\mu(\slashed{P}-\slashed{Q}+m_e)\gamma^\nu}{-2P\cdot Q+Q^2+i\epsilon}+(Q\leftrightarrow-Q)\right].
\end{equation}
In contracting with the tensor $Q^\mu Q^\nu-Q^2 g^{\mu\nu}$, we obtain
\begin{equation}
    F(K=0)=i\int \frac{d^4Q}{(2\pi)^4}\frac{16\pi^2\alpha T(Q^2)}{Q^4}\left[\frac{Q^2\slashed{P}+2(P\cdot Q-Q^2)\slashed{Q}-3m_e Q^2}{-2P\cdot Q+Q^2+i\epsilon}+(Q\leftrightarrow-Q)\right].
\end{equation}
Since we are only interested in the action of this operator on the spinor of an electron on shell, we may replace $\slashed{P}\to m_e$. Furthermore, the integral of $\int Q^\mu F(Q^2,P\cdot Q) d^4 Q\propto P^\mu$, so we may effectively replace $\slashed{Q}\to (Q\cdot P)\slashed{P}/m_e^2\to (Q\cdot P)/m_e$. Thus, we finally obtain
\begin{equation}
    F(K=0)=i\int \frac{d^4Q}{(2\pi)^4}\frac{16\pi^2\alpha T(Q^2)}{Q^2 m_e}\left[\frac{2m_e^2+P\cdot Q}{2P\cdot Q-Q^2-i\epsilon}-\frac{2m_e^2-P\cdot Q}{2P\cdot Q+Q^2+i\epsilon}\right].
\end{equation}
\end{widetext}
The integral is in a form that can be performed explicitly. However, we do not actually need to. It suffices to note that, differently from the nucleon case, the momentum flowing through the electron propagator cannot be much larger than $Q^2\gg m_e^2$, otherwise the electron propagator strongly suppresses the integration. Therefore, the order of magnitude of $T(Q^2\ll m_\mu^2)=\alpha g_{\phi \mu}/(6\pi^2 m_\mu)$ shows immediately that $F(K=0)\sim \alpha^2 g_{\phi\mu} m_e/m_\mu$. Thus, compared with the nucleon coupling, the coupling to electrons is parametrically suppressed by the small ratio $m_e/m_\mu \sim 0.01$. We therefore neglect the effective coupling to electrons.

For vector bosons, the triangle diagram vanishes due to charge conjugation invariance. The direct kinetic mixing between the vector boson and the photon does not produce fifth forces on neutral bodies, since the proton and the electron couple with opposite charges. 
More generally, higher-order diagrams with multiple photon insertions must respect electromagnetic gauge invariance. As a result, in the soft limit they cannot generate an unsuppressed coupling to the charge monopole of a neutral body: the electromagnetic current vanishes at zero momentum transfer, and the remaining gauge-invariant response is encoded in operators suppressed by powers of the exchanged momentum. Therefore, no long-range monopole fifth force is induced.

\section{E. Muon depletion by long-range vector force}

For sufficiently light vectors or sufficiently large vector couplings $g_{V\!\mu}$, the repulsive interaction among muons generates a potential $U_\mu$ that can substantially reduce their equilibrium abundance. In other words, the muon population sourcing the new force is itself depleted by the force it generates. This feedback also limits the associated vector pressure $P_V$, which never becomes important compared with gravitational pressure. For the lightest mediators, depletion drives muons into the nondegenerate regime before vector cooling becomes important, eliminating the NS cooling bound in Fig.~1 of the main text.

To quantify these effects, we first estimate the potential $U_\mu$. Since the force range is much smaller than the NS radius (see the upper axis of Fig.~1), we work in the local-density limit $m_V R_{\rm NS}\gg1$. The muonic vector field $V^\mu$ is described by the Lagrangian  
\begin{equation}
    \mathcal{L}\supset\bar\mu(i\slashed{\partial}-m_\mu)\mu-g_{V\!\mu}V_\alpha\bar\mu\gamma^\alpha\mu
    -\frac{1}{4}F_{\alpha\beta}F^{\alpha\beta}+\frac{1}{2} m_V^2V_\alpha V^\alpha\,,
\end{equation}
where $F_{\alpha\beta}=\partial_\alpha V_\beta-\partial_\beta V_\alpha$ and $m_V$ is the vector mass. The equation of motion is the Proca equation
\begin{equation}
    \partial_\beta F^{\beta\alpha}+m_V^2V^\alpha=g_{V\!\mu}j^\alpha\,,
\end{equation}
where $j^\alpha=\bar\mu\gamma^\alpha\mu$ is the muon current. For a static medium at rest, time derivatives vanish and spatial gradients can be neglected in the local-density limit. Only the component $V^0$ survives so that $m_V^2V^0=g_{V\!\mu}n_\mu$, where the muon density is $n_\mu=\bar\mu\gamma_0\mu$.

The vector field shifts the energy of every muon by $U_\mu=g_{V\!\mu} V^0$ or explicitly
\begin{equation}\label{eq:U}
U_\mu=\frac{g_{V\!\mu}^2 n_\mu}{m_V^2}.
\end{equation}
Thus, in the local-density limit, the new vector potential depends on  $g_{V\!\mu}$ and $m_V$ only through $|g_{V\!\mu}/m_V|$. The effective muon dispersion relation is
\begin{equation}\label{eq:muon-dispersion}
 E_{\bf p}=U_\mu+\sqrt{m_\mu^2+{\bf p}^2}.
\end{equation}
The associated vector-field energy density and pressure are equal, $\epsilon_V=P_V=U_\mu n_\mu/2$, giving
\begin{equation}
P_V =\frac{g_{V\!\mu}^2 n_\mu^2}{2m_V^2},
\end{equation}
which adds to the usual free-gas muon pressure.

For degenerate muons, $\mu_e=U_\mu+(m_\mu^2+p_{\rm F,\mu}^2)^{1/2}$ and therefore $p_{\rm F,\mu}^2=(\mu_e-U_\mu)^2-m_\mu^2$, with the physical solution satisfying $\mu_e-U_\mu\ge m_\mu$. With $n_\mu=p_{\rm F,\mu}^3/3\pi^2$ and Eq.~\eqref{eq:U}, the muon density is given by
\begin{equation}
  n_\mu=\frac{[(\mu_e-g_{V\!\mu}^2 n_\mu/m_V^2)^2-m_\mu^2]^{3/2}}{3\pi^2}.
\end{equation}
We express the muon abundance relative to the electron density as $y=n_\mu/n_e$. Taking the electrons to be ultrarelativistic, $n_e=\mu_e^3/3\pi^2$, we find
\begin{equation}\label{eq:y-equation}
  y=\left[\left(1-y \,\eta_V\right)^2-\left(m_\mu/\mu_e\right)^2\right]^{3/2},
\end{equation}
where the strength of the backreaction is controlled by the dimensionless parameter
\begin{equation}
  \eta_V=\frac{g_{V\!\mu}^2\mu_e^2}{3\pi^2m_V^2}.
\end{equation}
Since $y=\mathcal{O}(1)$ in the absence of the new force, backreaction becomes important parametrically for $\eta_V\agt 1$, corresponding to
\begin{equation}\label{eq:crit-coupling}
  g_{V\!\mu}\agt \frac{\sqrt3 \pi\, m_V}{\mu_e}.
\end{equation}
This is our estimate for the onset of muon depletion.

In principle, these expressions require determining $\mu_e$ self-consistently as muons are depleted, imposing charge neutrality, beta equilibrium, and the nuclear equation of state. In practice, the resulting correction is small for our parametric estimates. For our reference conditions (see End Matter),  $\mu_e=200~{\rm MeV}$ without muon depletion and rises to 234~MeV for complete depletion at fixed proton abundance, a modest 17\% increase. Therefore, the precise value of $\mu_e$ does not strongly affect our estimates.

For a numerical estimate of the onset of muon depletion  we use $\mu_e=200~{\rm MeV}$. In the absence of the new force, $m_\mu=105.66~{\rm MeV}$ gives $y=0.61$. The criterion $\eta_V\simeq1$ provides $g_{V\!\mu}\simeq2.7\times10^{-8}\,m_V/{\rm eV}$, corresponding to the lower oblique black line in Fig.~1.

In the strongly depleted limit, when $y\ll 1$, Eq.~\eqref{eq:y-equation} is solved by
\begin{equation}
  y\simeq \frac{1}{\eta_V}\left(1-\frac{m_\mu}{\mu_e}\right)=\frac{1}{\eta_V}\,\frac{\Delta_\mu}{\mu_e},
\end{equation}
where $\Delta_\mu=\mu_e-m_\mu$. Therefore
\begin{equation}\label{eq:DegMuonAbundance}
  n_\mu\simeq\frac{m_V^2\Delta_\mu}{g_{V\!\mu}^2}
  \quad\hbox{and}\quad
  P_V\simeq\frac{m_V^2\Delta_\mu^2}{2g_{V\mu}^2}. 
\end{equation}
Thus, in the strongly depleted regime, both $n_\mu$ and $P_V$ decrease with increasing $g_{V\!\mu}^2/m_V^2$. 

Since $P_V$ also vanishes as $g_{V\!\mu}^2/m_V^2\to0$, it reaches a maximum at intermediate depletion. Holding $\mu_e=200~{\rm MeV}$ fixed, we find a maximum of the new pressure contribution of $P_V^{\rm max}\simeq 0.2~{\rm MeV\,fm^{-3}}$. This is much smaller than the characteristic gravitational pressure $P_G=3 G M_{\rm NS}^2/8\pi R_{\rm NS}^4\simeq44\,{\rm MeV}{\rm fm}^{-3}$ for a NS of radius $R_{\rm NS}\simeq11\,{\rm km }$ and mass $M_{\rm NS}\simeq1.8\,{\rm M_{\odot}}$.

Muon depletion does not immediately quench vector cooling. While the initially dominant collective emission is suppressed as $n_\mu$ decreases, incoherent muon scattering becomes relatively more important. In the degenerate regime, this emissivity is nearly independent of $n_\mu$ because muon depletion is compensated by reduced degeneracy suppression. Thus, at fixed $m_V$, the emissivity continues to grow as $g_{V\!\mu}^2$. 

Eventually, however, muons become nondegenerate. This occurs when their kinetic Fermi energy approximately equals the NS temperature. Since $T\ll m_\mu$, they are then deeply nonrelativistic. Their kinetic Fermi energy is $(m_\mu^2+p_{{\rm F},\mu}^2)^{1/2}-m_\mu\simeq p_{{\rm F},\mu}^2/2m_\mu$ so that the transition occurs for $p_{{\rm F},\mu}^2/2m_\mu\simeq T$. 

In this regime, muons follow a Maxwell-Boltzmann distribution, and beta equilibrium still implies $\mu_\mu=\mu_e$, so that
\begin{equation}
n_\mu\simeq n_Q\,e^{(\Delta_\mu-U_\mu)/T},
\end{equation}
where $n_Q=2(m_\mu T/2\pi)^{3/2}$. Hence, the muon abundance satisfies 
\begin{equation}
n_\mu\simeq\frac{m_V^2}{g_{V\!\mu}^2}
\left[\Delta_\mu-T\log\left(\frac{n_\mu}{n_Q}\right)\right].
\end{equation}
Up to logarithmic corrections, it retains the scaling of Eq.~\eqref{eq:DegMuonAbundance} also in the Maxwell-Boltzmann regime. 

The logarithmic correction is very small, so that we may use $n_\mu=(m_V/g_{V\!\mu})^2\Delta_\mu=p_{{\rm F},\mu}^3/3\pi^2$. At the transition, $p_{{\rm F},\mu}^2/2m_\mu\simeq T$ so that $p_{{\rm F},\mu}\simeq (2m_\mu T)^{1/2}$. Thus the transition occurs for
\begin{equation}
  g_{V\!\mu}\simeq m_V\,\frac{\sqrt3 \pi (\mu_e-m_\mu)^{1/2}}{(2m_\mu T)^{3/4}}.
\end{equation}
Numerically, with the parameters inspired by our reference model, $T=8.9~{\rm keV}$ and $\mu_e=234~{\rm MeV}$ as the approximate muon-depleted value, the transition occurs for $g_{V\!\mu}\simeq3.8\times10^{-5}\,m_V/{\rm eV}$, corresponding to the upper oblique black line in Fig.~1.

For parameters above this line, the dominant emission is from incoherent collisions of nondegenerate muons with electrons and protons. For nondegenerate and nonrelativistic muons, the matrix element for incoherent scattering is
parametrically estimated as
\begin{equation}
\mathcal{M}_V^{(\mu f)}\sim g_{V\!\mu}
\frac{m_M}{\omega\,m_\mu}\mathcal{M}_{\mu f\to\mu f}
\end{equation}
with $f=e$ or $p$. No muon degeneracy factor suppresses the emission.  The energy-loss rate is then parametrically
\begin{equation}
    Q_{V}^{(\mu f)}\sim \frac{g_{V\!\mu}^2\alpha^2}{m_\mu^2}\,n_\mu\,n_f\,T.
\end{equation}
The scaling $n_\mu\propto m_V^2/g_{V\mu}^2$ therefore yields a rate parametrically independent of the muonic vector coupling.

As discussed in the main text, for $m_V<10^{-7}\,{\rm eV}$, muon depletion renders the muons nondegenerate at such small couplings that the new energy-loss rate never becomes competitive with standard neutrino cooling, eliminating the bounds (see Fig.~1).

A similar parametric estimate  applies to the hot proto-neutron star at the center of a core-collapse SN. Taking $T_{\rm SN}\sim30\,{\rm MeV}$ and $\Delta_\mu\simeq 130~{\rm MeV}$ as representative values, muons become nondegenerate for  $m_V\lesssim1.1\times10^{7}\,{\rm eV}\,\,g_{V\!\mu}$. The emission from muons then becomes parametrically independent of the coupling. Hence, SN~1987A cooling bounds on muonic vectors, at the level of $g_{V\!\mu}\simeq2\times10^{-9}$ for $m_V<T_{\rm SN}$, must weaken and eventually disappear for $m_V\lesssim2\times10^{-2}\,{\rm eV}$.

\end{document}